\documentclass[twocolumn,superscriptaddress,amsmath,amssymb,8pt]{revtex4-1}

\usepackage{graphicx}
\usepackage{color}
\usepackage[dvipsnames]{xcolor}
\usepackage{upgreek}
\usepackage{float}
\usepackage{upgreek}
\usepackage{amsmath,amssymb}
\usepackage{natmove}

\begin{document}

\title{Microscopic imaging of elastic deformation in diamond via in-situ stress tensor sensors}

\author{D. A. Broadway}
\affiliation{School of Physics, University of Melbourne, Parkville, VIC 3010, Australia}
\affiliation{Centre for Quantum Computation and Communication Technology, School of Physics, University of Melbourne, Parkville, VIC 3010, Australia}

\author{B. C. Johnson}
\affiliation{School of Physics, University of Melbourne, Parkville, VIC 3010, Australia}
\affiliation{Centre for Quantum Computation and Communication Technology, School of Physics, University of Melbourne, Parkville, VIC 3010, Australia}

\author{M. S. J. Barson}
\affiliation{Laser Physics Centre, Research School of Physics and Engineering, Australian National University, Canberra, ACT 2601, Australia}

\author{S. E. Lillie}
\affiliation{School of Physics, University of Melbourne, Parkville, VIC 3010, Australia}
\affiliation{Centre for Quantum Computation and Communication Technology, School of Physics, University of Melbourne, Parkville, VIC 3010, Australia}

\author{N. Dontschuk}
\affiliation{School of Physics, University of Melbourne, Parkville, VIC 3010, Australia}
\affiliation{Centre for Quantum Computation and Communication Technology, School of Physics, University of Melbourne, Parkville, VIC 3010, Australia}

\author{D. J. McCloskey}
\affiliation{School of Physics, University of Melbourne, Parkville, VIC 3010, Australia}

\author{A. Tsai}
\affiliation{School of Physics, University of Melbourne, Parkville, VIC 3010, Australia}

\author{T. Teraji}
\affiliation{National Institute for Materials Science, Tsukuba, Ibaraki 305-0044, Japan}

\author{D. A. Simpson}
\affiliation{School of Physics, University of Melbourne, Parkville, VIC 3010, Australia}

\author{A. Stacey}
\affiliation{Centre for Quantum Computation and Communication Technology, School of Physics, University of Melbourne, Parkville, VIC 3010, Australia}
\affiliation{Melbourne Centre for Nanofabrication, Clayton, VIC 3168, Australia}

\author{J. C. McCallum}
\affiliation{School of Physics, University of Melbourne, Parkville, VIC 3010, Australia}

\author{J. E. Bradby}
\affiliation{Electronic Materials Engineering, Research School of Physics and Engineering, Australian National University, Canberra, ACT 2601, Australia}

\author{M. W. Doherty}
\affiliation{Laser Physics Centre, Research School of Physics and Engineering, Australian National University, Canberra, ACT 2601, Australia}

\author{L. C. L. Hollenberg}
\email{lloydch@unimelb.edu.au}
\affiliation{School of Physics, University of Melbourne, Parkville, VIC 3010, Australia}
\affiliation{Centre for Quantum Computation and Communication Technology, School of Physics, University of Melbourne, Parkville, VIC 3010, Australia}	

\author{J.-P. Tetienne}
\email{jtetienne@unimelb.edu.au}
\affiliation{School of Physics, University of Melbourne, Parkville, VIC 3010, Australia}

\begin{abstract}
The precise measurement of mechanical stress at the nanoscale is of fundamental and technological importance. In principle, all six independent variables of the stress tensor, which describe the direction and magnitude of compression/tension and shear stress in a solid, can be exploited to tune or enhance the properties of materials and devices. However, existing techniques to probe the local stress are generally incapable of measuring the entire stress tensor. 
Here, we make use of an ensemble of atomic-sized in-situ strain sensors in diamond (nitrogen-vacancy defects) to achieve spatial mapping of the full stress tensor, with a sub-micrometer spatial resolution and a sensitivity of the order of 1 MPa (corresponding to a strain of less than $10^{-6}$). To illustrate the effectiveness and versatility of the technique, we apply it to a broad range of experimental situations, including mapping the elastic stress induced by localized implantation damage, nano-indents and scratches. In addition, we observe surprisingly large stress contributions from functional electronic devices fabricated on the diamond, and also demonstrate sensitivity to deformations of materials in contact with the diamond. 
Our technique could enable in-situ measurements of the mechanical response of diamond nanostructures under various stimuli, with potential applications in strain engineering for diamond-based quantum technologies and in nanomechanical sensing for on-chip mass spectroscopy.
\end{abstract}

\maketitle

Pressure is a powerful thermodynamic variable often used to modify a material's properties~\cite{Li2014}. 
Most notably, strain enhances the charge carrier mobility in modern electronics~\cite{Chu2009,Shin2016,Adams2011}. It allows the tuning of the optical properties of materials~\cite{Smith2008,Ni2013,Jacobsen2006}, can confer them with a ferroelectric nature~\cite{Tang2015,Zhang2018}, or even make them better superconductors~\cite{Locquet1998}. Despite these successes, strain engineering is still a largely unexplored field considering the huge parameter space available; indeed, stress is characterized by six parameters (three axial components and three shear components, defining the stress tensor) which, in principle, can be continuously and independently varied over many orders of magnitude to optimize the functional properties of materials~\cite{Li2014}. 

Key to further innovations is the ability to characterise stress at the nanoscale, and in particular to quantitatively determine the six components of the stress tensor.  
Existing methods to probe stress in solids typically rely on the interaction between a beam of probe particles (usually electrons or photons) with the stressed material~\cite{Vanhellemont1993,Jain1996,Voloshin1983,Tang1994,Wright2011,Kato2012,Holt2013,Hytch2014,Whiteley2018}. However, these techniques are generally sensitive to only one or a convolution of the stress components. Exceptions include off-axis electron holography~\cite{Hytch2008}, but at the cost of sample destruction to produce suitably thin lamellae (under 200 nm thickness, which may cause strain release), and off-axis Raman spectroscopy~\cite{Loechelt1999}, but at the expense of a spatial resolution limited to millimeter scales. 

\begin{figure*}[t!]
	\centering
	\includegraphics[width=\textwidth]{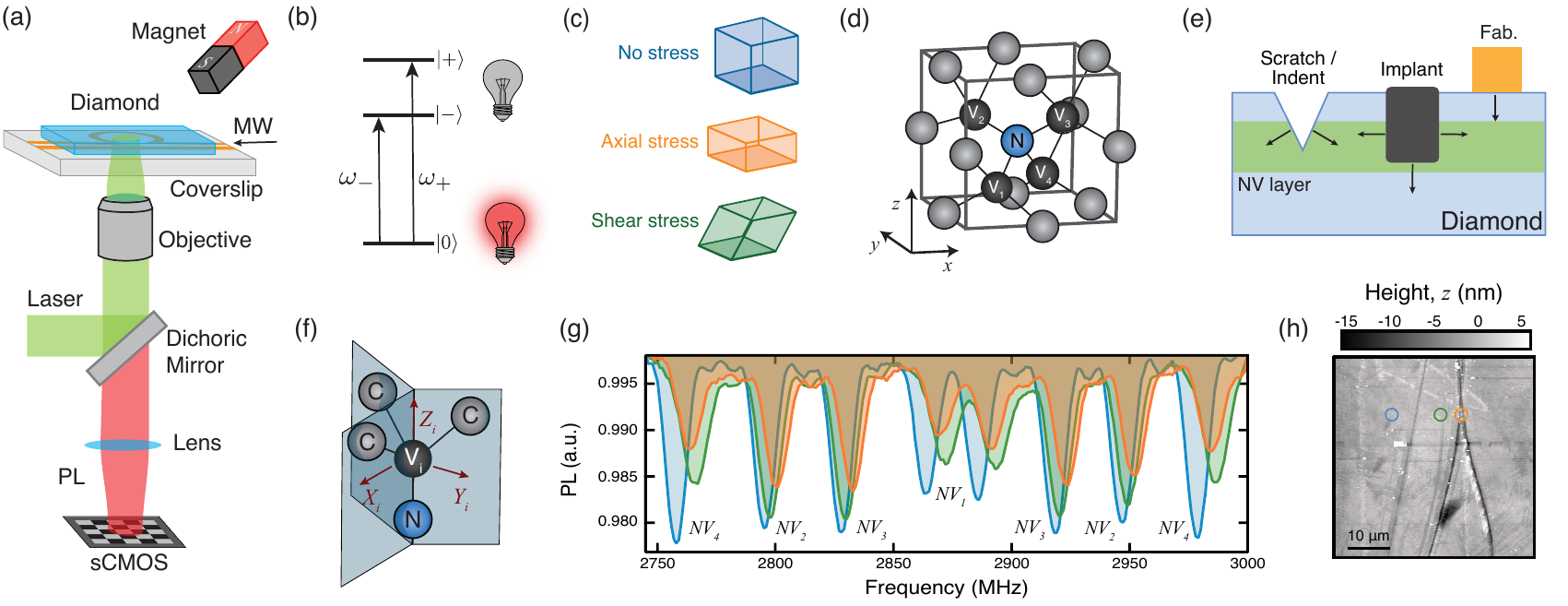}
	\caption{Stress-tensor mapping with nitrogen-vacancy centres. 
		(a) Diagram of the experimental set-up, depicting a diamond sensing chip mounted on a glass coverslip with a microwave (MW) resonator. The layer of nitrogen-vacancy (NV) centres in the diamond is illuminated by a green laser and imaging of the red NV photoluminescence (PL) is captured with a sCMOS camera.  
		(b) Fine structure of the NV electronic ground state showing the two spin transitions with frequencies $\omega_\pm$ that can be probed experimentally by virtue of spin-dependent PL (symbolized by light bulbs of different brightness).
		(c) Schematic of the unit cell of the diamond crystal under different stress conditions: no stress (blue), axial stress (orange), and shear stress (green).
		(d) Unit cell with the four NV orientations depicted as different vacancy locations V$_i$ ($i=1,2,3,4$). The stress tensor is expressed in the natural coordinate system of the cubic lattice, $(xyz)$.
		(e) Schematic cross-section of the diamond illustrating how the NV layer (green band) is used to probe the stress produced by local features.
		(f) Defect structure of the NV and its native coordinate system $(X_iY_iZ_i)$.  
		(g) Optically-detected magnetic resonance spectra under an external magnetic field of strength $B\sim50$~G, recorded from NV centres at three different locations in a scratched diamond indicated by circles in (h): away from scratch damage (blue), underneath a scratch (orange), and adjacent to a scratch (green). Each resonance is labelled according to its corresponding NV orientation. 
		(h) Atomic force microscopy (AFM) image of the diamond surface showing scratches created with a diamond scribe.
		 }
	\label{Fig: intro}
\end{figure*}

In this work, we describe a radically different approach, which relies on atomic-sized strain sensors embedded into the material to characterise the local strain. In recent years, several materials have been found to host such in-situ strain sensors, in the form of optically addressable point defects that exhibit strain-dependent energy levels. These include diamond~\cite{Grazioso2013,Doherty2014a,Barson2017}, silicon carbide~\cite{Falk2014} and silicon~\cite{Zhang2018}. To date, however, real-space imaging of the full stress tensor using in-situ sensors has remained elusive, in part due to the difficulty of addressing a sufficiently dense ensemble of sensors while retaining the capability to extract all the tensor components. In the case of the nitrogen-vacancy (NV) defect in diamond~\cite{Doherty2013}, employed in this work, previous works have achieved two-dimensional (2D) imaging of a single strain component~\cite{Trusheim2016,Bauch2018} and outlined a method to extract the entire tensor~\cite{Grazioso2013}. Single NV defects have also been proposed for nanomechanical sensing, including vector force sensing and mass spectroscopy~\cite{Barson2017}. Here we demonstrate wide-field 2D mapping of the full stress tensor with sub-micrometer spatial resolution, at room temperature. 
We apply the technique to spatially non-uniform strain features induced by scratches, implantation damage, nano-indents and devices fabricated on diamond. 
These results establish NV-based wide-field stress mapping as a powerful technique for nanoscale mechanical studies of diamond, which is an interesting material for nanomechanical applications with extreme mechanical properties and sometimes unexpected strain-stress relationships~\cite{Banerjee2018}. The technique may also facilitate strain engineering for diamond-based quantum applications~\cite{Macquarrie2013,Teissier2014,Ovartchaiyapong2014a,Barfuss2015,Chen2018,Macquarrie2015,Sohn2018}, and find applications in multiplexed nanomechanical sensing for mass spectroscopy and microfluidics~\cite{Gil-Santos2010,Arlett2011,Calleja2012,Hanay2015}. 

\section*{Results and discussion}


The experimental set-up is depicted in Figure~\ref{Fig: intro}a. It consists of a wide-field fluorescence microscope equipped with microwave excitation to allow the spin transition frequencies of the NV centres ($\omega_\pm$, Figure~\ref{Fig: intro}b) to be probed via optically-detected magnetic resonance spectroscopy, thanks to their spin-dependent photoluminescence (PL)~\cite{Rondin2014}. These frequencies depend on the spatial distribution of the unpaired spin density and so are sensitive to the local strain in the lattice, or equivalently~\cite{Barson2017} to the local stress which can have axial and shear components (Figure~\ref{Fig: intro}c). As we will see, the use of multiple NV orientations (Figure~\ref{Fig: intro}d) enables the full reconstruction of the local stress tensor. By using a thin layer of NV centres near the diamond surface, we can then spatially map the stress tensor in two dimensions and study the effect of localized strain-inducing features (Figure~\ref{Fig: intro}e). We note that the presence of the NV centres and other unavoidable defects such as substitutional nitrogen may also contribute to the net measured stress~\cite{Biktagirov2017}.    

The fine structure of the NV electronic ground state (Figure~\ref{Fig: intro}b) is governed by the spin Hamiltonian~\cite{Doherty2012},
\begin{eqnarray} \label{eq:Ham}
\begin{aligned}
H_i &= (D + \mathcal{M}_{Z_i})S_{Z_i}^2 + \gamma_{\rm NV}\vec{B}\cdot\vec{S}  \\
&\quad - \mathcal{M}_{X_i} ( S_{X_i}^2 - S_{Y_i}^2)  + \mathcal{M}_{Y_i}( S_{X_i}S_{Y_i} + S_{Y_i}S_{X_i}),
\end{aligned}
\end{eqnarray} 
where $D \approx 2.87$~GHz is the temperature-dependent zero-field splitting parameter, $\gamma_{\rm NV}=28.035(3)$~GHz T$^{-1}$ is the NV gyromagnetic ratio~\cite{Doherty2013}, $\vec{S}_i=(S_{X_i},S_{Y_i},S_{Z_i})$ are the spin-1 operator, $\vec{B}$ is the applied magnetic field, and $\vec{\cal M}_i=({\cal M}_{X_i},{\cal M}_{Y_i},{\cal M}_{Z_i})$ is the effective electric field associated with mechanical stress, where we neglect any residual true electric field caused by charge effects~\cite{Dolde2011,Broadway2018} (Supporting Information, Section IV). Here $(X_iY_iZ_i)$ is the coordinate system of the NV defect structure (Figure~\ref{Fig: intro}f), and the index $i=1,2,3,4$ denotes the NV orientation with respect to the diamond unit cell (Figure~\ref{Fig: intro}d). The relation between $\vec{\cal M}_i$ and the local stress is most conveniently expressed by defining the stress tensor, $\overleftrightarrow\sigma$, with respect to the diamond unit cell coordinate system $(xyz)$ defined in Figure~\ref{Fig: intro}d~\cite{Barson2017,Barfuss2018}. This gives
\begin{eqnarray} \label{eq:M}
\mathcal{M}_{X_i} &=& b\Sigma_X^{\rm axial} + c\Sigma_{X_i}^{{\rm shear}} \nonumber \\
\mathcal{M}_{Y_i} &=& \sqrt{3}b\Sigma_Y^{\rm axial} + \sqrt{3}c\Sigma_{Y_i}^{{\rm shear}} \\
\mathcal{M}_{Z_i} &=& a_1\Sigma_Z^{\rm axial}  + 2a_2\Sigma_{Z_i}^{{\rm shear}} \nonumber
\end{eqnarray}
where the stress susceptibility parameters are $a_1 = 4.86(2)$, $a_2 = -3.7(2)$, $b = -2.3(3)$, and $c = 3.5(3)$ in units of MHz GPa$^{-1}$~\cite{Barson2017}, and we introduced the quantities
\begin{eqnarray} \label{eq:axial}
\Sigma_X^{\rm axial} &=& -\sigma_{xx}-\sigma_{yy}+2\sigma_{zz} \nonumber \\
\Sigma_Y^{\rm axial} &=& \sigma_{xx} - \sigma_{yy} \\
\Sigma_Z^{\rm axial} &=& \sigma_{xx}+\sigma_{yy}+\sigma_{zz}, \nonumber
\end{eqnarray}
which capture the contribution of the axial stress components and are independent of the NV orientation~\cite{AliMomenzadeh2016}, and
\begin{eqnarray} \label{eq:shear}
\Sigma_{X_i}^{{\rm shear}} &=& 2f_i \sigma_{xy}+g_i\sigma_{xz}+f_ig_i\sigma_{yz} \nonumber \\
\Sigma_{Y_i}^{{\rm shear}} &=& g_i\sigma_{xz}-f_ig_i \sigma_{yz} \\
\Sigma_{Z_i}^{{\rm shear}} &=& f_i\sigma_{xy} -g_i \sigma_{xz} -f_ig_i \sigma_{yz}, \nonumber
\end{eqnarray}
which describe the effect of the shear stress components. In Equation~(\ref{eq:shear}), the functions $f_i$ and $g_i$ evaluate to $\pm1$ depending on the NV orientation. Namely, using the vacancy positions $i=1,2,3,4$ as defined in Figure~\ref{Fig: intro}d, we have $(f_1,f_2,f_3,f_4)=(+1,-1,-1,+1)$ and $(g_1,g_2,g_3,g_4)=(-1,+1,-1,+1)$. We note that the spin-mechanical interaction of the NV is invariant under inversion of the nitrogen and the vacancy, therefore there are only four orientations to consider in contrast to true electric fields~\cite{Doherty2014}. 

Because the stress tensor is described by six independent parameters, a measurement relying on a single NV centre is not sufficient to infer the full stress tensor even if the effective field $\vec{\cal M}_i$ is completely determined~\cite{Grazioso2013}. However, by using a small ensemble of NV centres with multiple orientations, it is possible to determine all six stress components in the corresponding volume, assuming a uniform stress within this volume. To see that, we consider the limit of small magnetic fields $|\vec{B}|\ll D$, for which the spin transition frequencies are given by~\cite{Dolde2011,Barson2017}
\begin{eqnarray} \label{eq:omega}
(\omega_\pm)_i&=&D+M_{Z_i}  \\
& & \pm\sqrt{(\gamma_{\rm NV}B_{Z_i})^2+(M_{X_i})^2+(M_{Y_i})^2}. \nonumber
\end{eqnarray} 
By using Equation~(\ref{eq:M}-\ref{eq:shear}), we find that the sum frequencies $(\omega_++\omega_-)_i=D+M_{Z_i}$ of the four possible orientations can be used to uniquely determine the three shear stress components $(\sigma_{xy},\sigma_{xz},\sigma_{yz})$ as well as the sum of the axial components, $(\sigma_{xx}+\sigma_{yy}+\sigma_{zz})$. The axial components are then determined individually by using the difference frequencies $(\omega_+-\omega_-)_i=2\sqrt{(\gamma_{\rm NV}B_{Z_i})^2+(M_{X_i})^2+(M_{Y_i})^2}$, which in general is an over-determined problem when the magnetic field $\vec{B}$ is known. This is the basis of our method to determine the full stress tensor (see further details in Supporting Information, Section I and II). 

Experimentally, we measured the frequencies $(\omega_\pm)_i$ for the four possible NV orientations by recording optically-detected magnetic resonance (ODMR) spectra under a small applied magnetic field $\vec{B}$ aligned so that all eight transitions can be resolved simultaneously~\cite{Tetienne2017,Broadway2018}, as shown in Figure~\ref{Fig: intro}g. As a preliminary test, we investigated the stress induced from scratching the diamond with a diamond tipped scribe~\cite{Lillie2018}, generating cuts that are less than a micron wide and range from 5 to 20~nm in depth (Figure~\ref{Fig: intro}h), with the NV layer extending from about 5 to 30 nm below the surface~\cite{Tetienne2018}. A reference ODMR spectrum (i.e. away from any scratch) is shown in blue in Figure~\ref{Fig: intro}g, which is used to infer the magnetic field, here $(B_x,B_y,B_z)\approx(22,-13,-40)$~G. An ODMR spectrum taken from underneath the scratch is shown in orange and exhibits a shift of about 10 MHz that is relatively uniform across all the resonances, suggesting a stress field in the GPa range that is dominated by axial stress components. An ODMR spectrum taken from a region adjacent to the scratch is shown in green, and exhibits shifts that are markedly different for different NV orientations. This is the signature of a large contribution from shear stress. 

\begin{figure*}[t!]
	\includegraphics[width=0.9\textwidth]{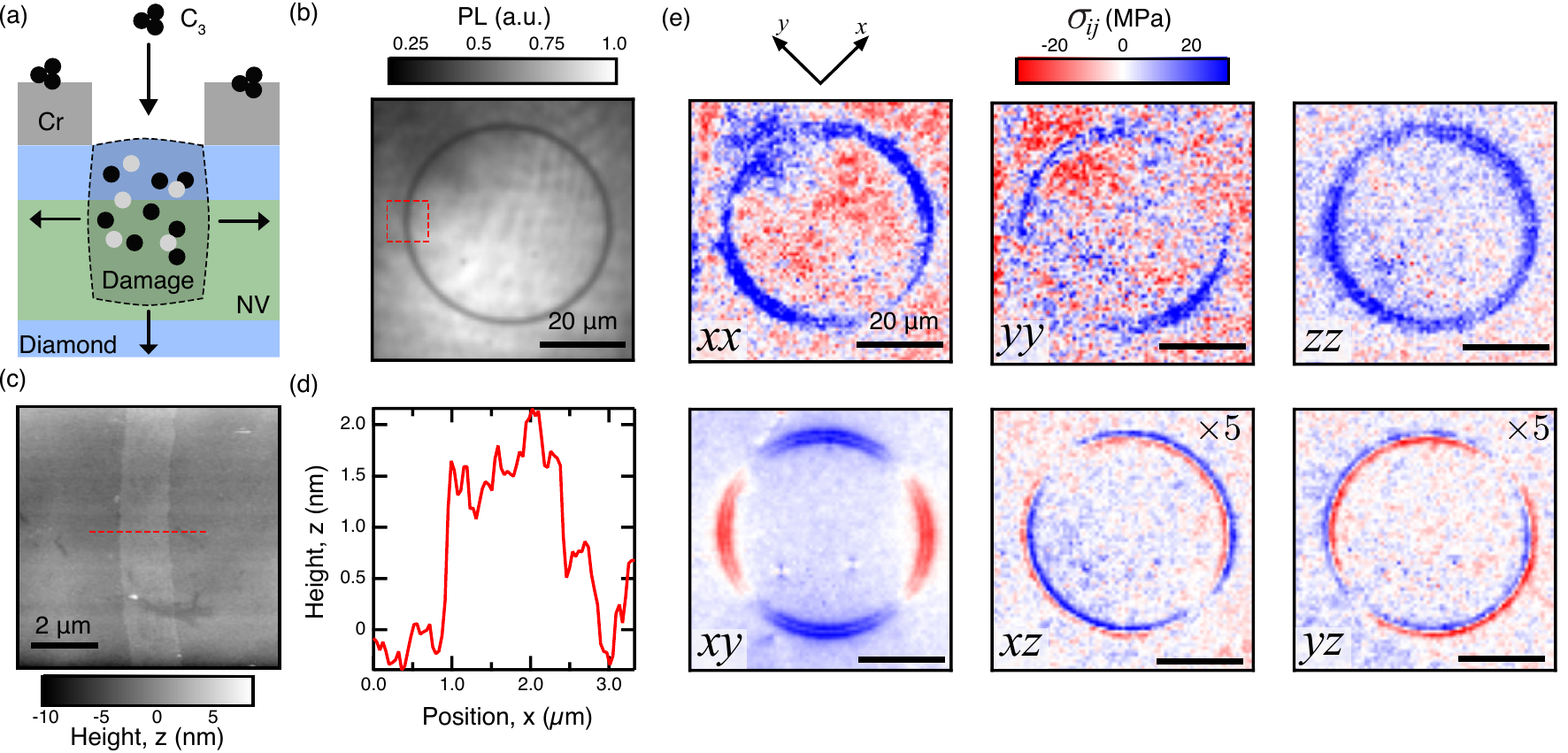}
	\caption{Stress induced by implantation damage. 
		(a) Schematic of the stress induced by locally implanting C$_3$ molecules in the diamond. 
		(b) NV PL image of an annulus-shaped implanted region.
		(c) AFM image of a segment of the annulus (dashed box in (b)). 
		(d) AFM profile across the implanted region (dashed line in (c)). 	
		(e) Spatial maps of the six stress tensor components measured with the NV sensors in the same region as in (b). The amplitude of the $\sigma_{xz}$ and $\sigma_{yz}$ components has been multiplied by a factor of 5. 
	}
	\label{Fig: implantation}
\end{figure*}


To test our method for reconstructing the full stress tensor, we first implanted C$_3$ molecules into the diamond to create localized regions of damage extending 5-10 nm below the diamond surface. Implantation is commonly used in diamond to introduce dopants for electrical devices~\cite{Tsubouchi2006} or produce buried graphitic electrical wires~\cite{Forneris2015}. The resultant confined amorphous carbon has a different density to diamond thus introducing an embedded force that pushes in all directions (Figure~\ref{Fig: implantation}a). It also causes a reduction in PL from the NV defects by about 25\%, as seen in Figure~\ref{Fig: implantation}b. Here, the implant pattern is a circular ring of $1.5~\mu$m width and $50~\mu$m diameter. The embedded force from the damage is sufficient to cause a bulging of the diamond surface~\cite{Olivero2013}, which was measured by atomic force microscopy (AFM) to be on the order of 2~nm in this case (Figure~\ref{Fig: implantation}c,d). 

The spatial maps of the six stress tensor components near this circular ring are shown in Figure~\ref{Fig: implantation}e, revealing ring-shaped patterns with stress values of up to tens of MPa (corresponding to ODMR frequency shifts in the 100 kHz range), much smaller than for the scratches discussed above. The pixel-to-pixel noise is about 1~MPa (standard deviation) for the three shear stress components, against 10 MPa for the axial stress components, with a total acquisition time of about 10 hours. This difference originates from the presence of the magnetic field projection $B_{Z_i}$ in the last term of Equation~(\ref{eq:omega}) used to separate the individual axial components, which results in a reduced sensitivity when $(\gamma_{\rm NV}B_{Z_i})^2\gg(M_{X_i})^2+(M_{Y_i})^2$ (Supporting Information, Section III). This loss of sensitivity could be mitigated by performing sequential measurements minimising $B_{Z_i}$ for each NV orientation. Nevertheless, all stress components exhibit resolvable features, and the different symmetries indicate they have been sensibly separated. We note that the axial components have a positive value near the damaged region indicating a compressive stress, consistent with an expansion of the lattice caused by the implantation~\cite{Olivero2013}. The smallest spatial features in Figure~\ref{Fig: implantation}e have a size of the order of $1~\mu$m. Moreover, in the Supporting Information (Figure S2) we show that implanted regions separated by less than $1~\mu$m can be resolved in the stress maps, indicating a sub-$\mu$m spatial resolution close to the optical diffraction limit ($\approx400$~nm~\cite{Simpson2016}).

The values of stress measured here are consistent with Raman spectroscopy~\cite{Olivero2013} performed on a similar sample (Supporting Information, Figure S4), which indicates an axial stress of $7\pm2$~MPa in the damaged region. This validates the NV technique as a sensitive and quantitative tool to characterise the stress in diamond, with the added benefit of access to the complete tensor information.



\begin{figure*}[tb!]
	\includegraphics[width=0.9\textwidth]{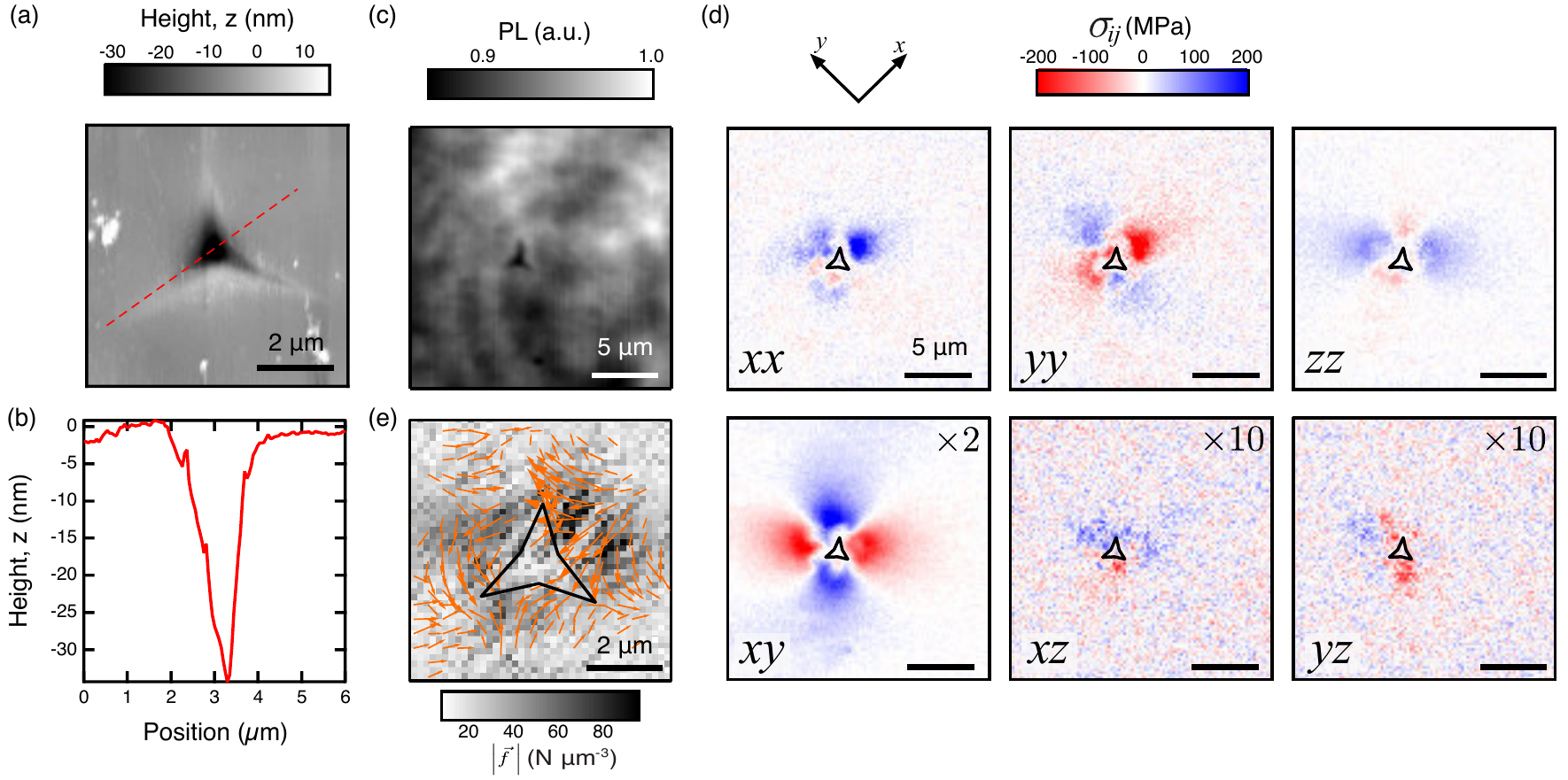}
	\caption{Stress induced by a nano-indent. 
		(a) AFM image of a representative nano-indent in the diamond surface.
		(b) AFM profile across the indent (dashed line in (a)).
		(c) NV PL image of a region containing a nano-indent.
		(d) Spatial maps of the six stress tensor components in the same region as in (b). The black shapes represent the contour of the nano-indent as determined from the PL image. The amplitude of the $\sigma_{xz}$ and $\sigma_{yz}$ ($\sigma_{xy}$) components has been multiplied by a factor of 10 (2).  
		(e) Map of the body force deduced from the stress tensor using Equation~(\ref{Eq: force}). The colour denotes the magnitude $|\vec{f}|$ while the overlaid arrows indicate the projected force in the $xy$ plane.  
	}
	\label{Fig:indent}
\end{figure*}


As a second test structure, we performed a series of nano-indents in the diamond surface using a Berkovich tip with a tip radius of 200 nm to a load of 1 N, resulting in pyramidal impressions typically with a maximum depth of 30 nm (see AFM image of a representative indent in Figure~\ref{Fig:indent}a, and line cut in Figure~\ref{Fig:indent}b). Slight swelling around the perimeter of the impression but no major structural fractures are observed. Likewise, no discontinuities indicative of fracture formation were observed in the load-unload curve (Supporting Information, Figure S6). This suggests that the diamond has been plastically deformed under these indentation conditions similar to previous reports~\cite{Brookes1970,Gogotsi1998}. It is likely that the indented region contains pressure induced graphitic material as observed recently~\cite{Dub2017}. Figure~\ref{Fig:indent}c reveals some quenching of the NV PL within the impression (by about 10\%) due to this reduction in crystal quality. Beyond the perimeter of the indent the impact on the strain can be quantified. The stress maps are presented in Figure~\ref{Fig:indent}d and reveal stress values up to 0.2 GPa for the in-plane components ($\sigma_{xx},\sigma_{yy},\sigma_{xy}$), the other components being much weaker indicating that the deformed material pushes predominantly in the directions parallel to the surface of the undeformed diamond. 

To facilitate the interpretation of the stress maps, it is convenient to plot the body force instead, where the Cartesian components are derived from the stress tensor via
\begin{align}
f_j & = -\sum_i \partial_i \sigma_{ji} \label{Eq: force}
\end{align}
with $i,j=x,y,z$ ($\partial_i$ denotes the partial derivative with respect to Cartesian coordinate $i$). The $x$ and $y$ derivatives can be readily computed from the stress maps while the $z$ derivatives can be estimated through a suitable approximation (Supporting Information, Section V). The body force near the indent is shown in Figure~\ref{Fig:indent}e, revealing a complicated pattern. This pattern indicates that the tips of the impression push mostly outwards while there is an inward force at its flat edges. This illustrates that Berkovich indentation results in a significant shear stress component which is known to aid phase transformation~\cite{Gogotsi1998}.


\begin{figure}
	\includegraphics[width=0.85\columnwidth]{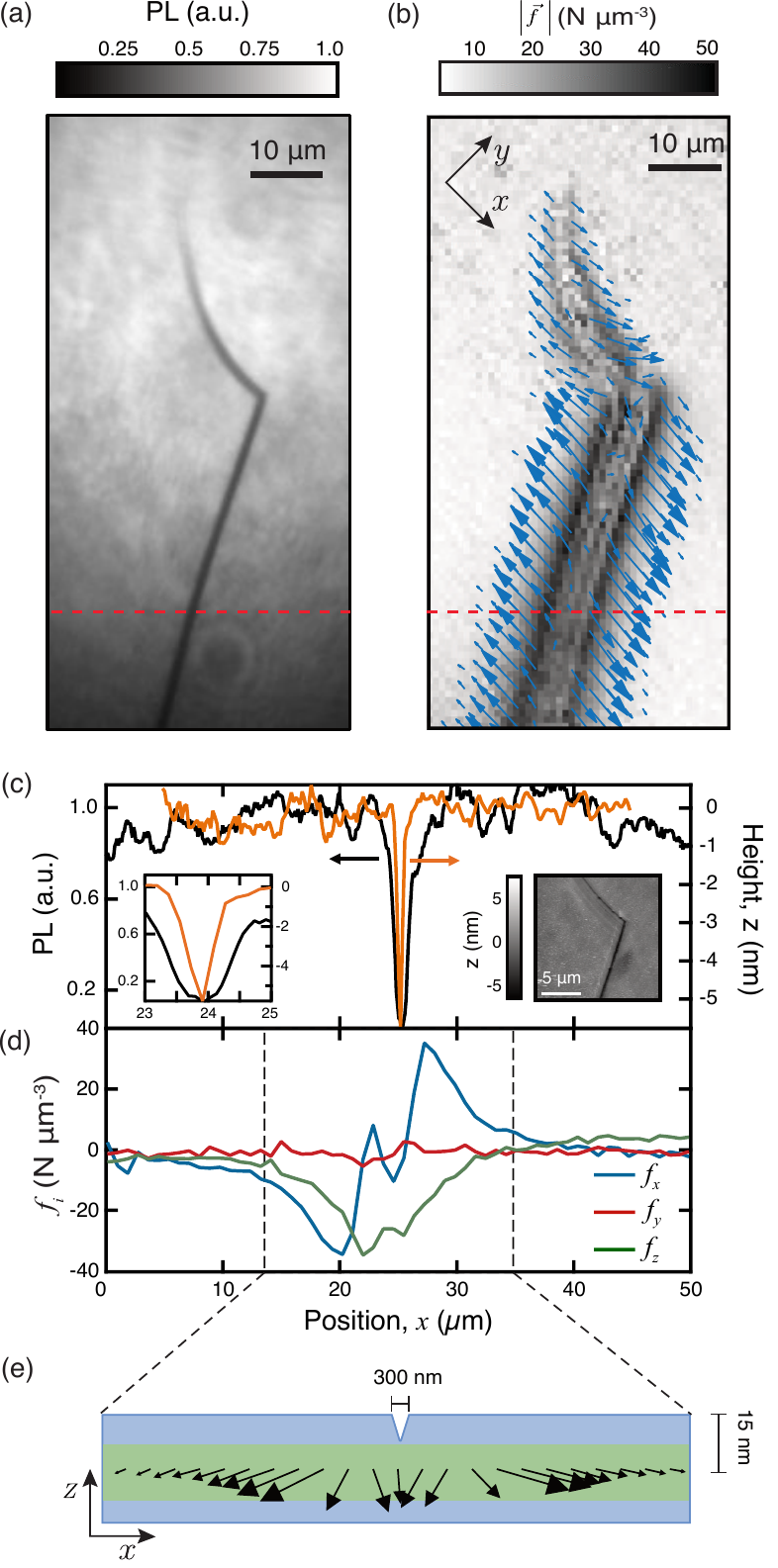}
	\caption{Stress induced by a superficial scratch. 
		(a) PL image of a region containing a scratch. 
		(b) Map of the body force magnitude for the region imaged in (a), with the overlaid arrows indicating the force projected in the $xy$ plane.   
		(c) Line cuts of the PL (black) and AFM profile (orange) taken along the red dashed line shown in (a,b). The left inset is a zoom-in; the right inset is an AFM image of the angled section of the scratch. 
		(d) Corresponding line cuts of the three body force components (in and out of plane).
		(e) Schematic of the diamond cross-section (not to scale), with the overlaid arrows indicating the body force projected in the $xz$ plane as derived from (d).}
	\label{Fig:scratch}
\end{figure}

We now return to scratches made with a diamond scribe to map the full stress tensor. Scratching the diamond surface can result in plastic deformation~\cite{Gogotsi1998}. Similar to the indent case, there is so sign of fracture of the diamond. A PL image of a scratched region is shown in Figure~\ref{Fig:scratch}a, where the scratch appears as a dark streak due to the reduction in crystal quality. The body force derived from the measured stress tensor (Supporting Information, Figure S7) is plotted in Figure~\ref{Fig:scratch}b. For this shallow scratch (5 nm deep at most, see AFM profile in Figure~\ref{Fig:scratch}c and image in inset), the magnitude of the body force reaches about 40~N $\mu$m$^{-3}$ corresponding to a stress of up to 0.2 GPa. Deeper scratches like those depicted in Figure~\ref{Fig: intro}g,h can produce in excess of 1 GPa of stress (Supporting Information, Figure S7). The body force decays from the center of the cut over distances of several micrometers, much larger than the width of the physical cut ($\approx300$~nm FWHM, see orange line in Figure~\ref{Fig:scratch}c) and the width of the PL quenching feature ($\approx1~\mu$m FWHM, black line in Figure~\ref{Fig:scratch}c). This illustrates that even a very shallow cut can elastically deform the diamond over distances significantly larger than the size of the cut itself. To gain more insight into the direction of the force, we plot the three force components across the scratch in Figure~\ref{Fig:scratch}d, represented as arrows overlaid on the diamond cross-section in Figure~\ref{Fig:scratch}e. Interestingly, the force remains mostly perpendicular to the diamond surface over several micrometers about the cut even though the NV sensors are located at a distance of $\approx15$~nm from the surface. This behaviour suggests a highly anisotropic propagation of the stress along the diamond surface.


\begin{figure*}
	\includegraphics[width=0.9\linewidth]{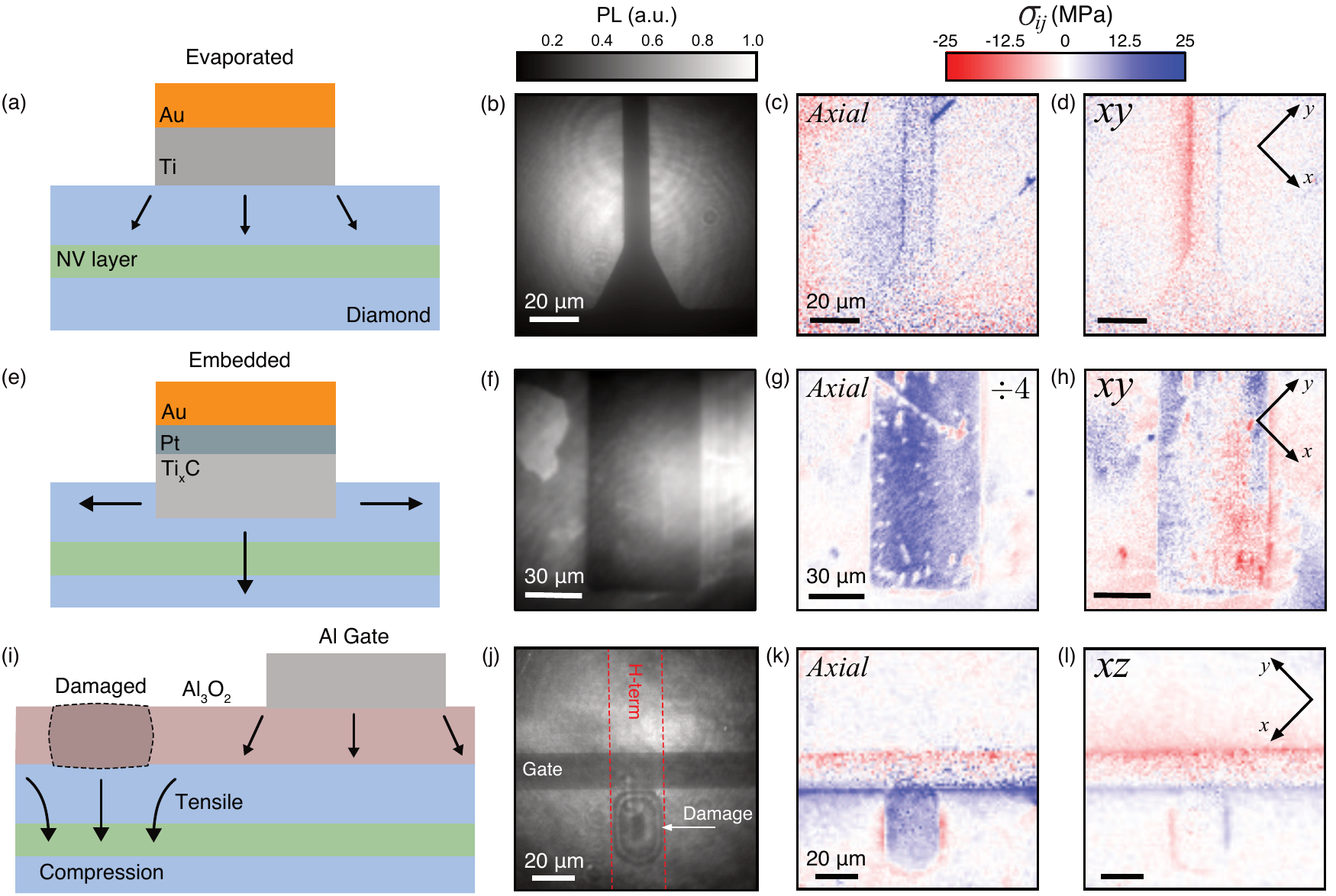}
	\caption{Stress induced by device fabrication. 
		(a) Schematic cross-section of a device consisting of an evaporated metallic wire (Ti/Au).
		(b) PL image of a typical device.
		(c,d) Corresponding maps of the total axial stress $\Sigma_Z^{\rm axial}=\sigma_{xx}+\sigma_{yy}+\sigma_{zz}$ (c) and of the shear stress component $\sigma_{xy}$.
		(e-h) Same as (a-d) but for an embedded TiC/Pt/Au contact.
		(i-l) Same as (a-d) but for a transistor device where the conductive channel is formed by a hydrogen-terminated (`H-term.') region of the diamond surface (delimited by red dashed lines in (j) but giving no PL contrast), insulated via a Al$_2$O$_3$ oxide layer and controlled by a top Al gate (appears dark in the PL). A defect formed in the oxide layer above the conductive channel while operating the device, visible through a fringe pattern in the PL. (l) shows the $\sigma_{xz}$ map. All the stress maps share the same color bar ranging from -25 to +25 MPa, except (g) where the stress values have been divided by a factor of 4 (i.e. ranging from -100 to +100 MPa).  
	}
	\label{Fig: Fab}
\end{figure*}

As a final illustration of the technique, we mapped the stress resulting from the fabrication of electronic devices on diamond, which is of interest for high-power and high-frequency electronics applications~\cite{Wort2008} and for studies of charge transport using NV-based magnetometry~\cite{Tetienne2017,Tetienne2018c}. First, we fabricated a strip of Ti/Au (thickness 10/100 nm) on the bare diamond surface by thermal evaporation and lift-off (Figure~\ref{Fig: Fab}a,b), and found that the strip induces a small but measurable compressive stress in the diamond (Figure~\ref{Fig: Fab}c,d) especially near the edges of the strip ($\sim10$~MPa). Here, we plot the sum of the axial stress components, $\Sigma_Z^{\rm axial}=\sigma_{xx}+\sigma_{yy}+\sigma_{zz}$, as this gives a significantly improved signal-to-noise ratio over the individual components (Supporting Information, Section III), as well as one of the shear stress components, $\sigma_{xy}$. We note that the weight of the deposited metal amounts to a pressure of about 0.02 Pa, and as such cannot account for the measured stress. Also visible in Figure~\ref{Fig: Fab}c are some polishing marks on the bare diamond surface, producing an axial stress of up to $\Sigma_Z^{\rm axial}\sim40$~MPa. 

Next, we fabricated TiC/Pt/Au contacts that extend about 15 nm into the diamond, through evaporation of a Ti/Pt/Au stack (thickness 10/10/70 nm) and subsequent annealing~\cite{Broadway2018} (Figure~\ref{Fig: Fab}e,f). Such embedded contacts are typically used to form low-resistance ohmic contacts with the two-dimensional hole gas (2DHG) present on a hydrogen-terminated diamond surface~\cite{Jingu2010,Broadway2018} as well as in high-power electronics~\cite{Hoshino2012}. As seen in the stress maps (Figure~\ref{Fig: Fab}g,h), there is a large compressive stress in the diamond below the TiC contact ($\Sigma_Z^{\rm axial}\sim100$~MPa). This is expected because TiC has a lower atomic density than diamond. It is relatively uniform across the contact except for some small spots (a few $\mu$m in size) where there is no stress, indicative of imperfections in the formed TiC layer.  

We then fabricated a transistor device based on the 2DHG at the diamond surface~\cite{Jingu2010,Hirama2007}, by first patterning the surface termination to form a conductive (hydrogen-terminated) channel on an otherwise insulating (oxygen-terminated) surface, covering the whole surface with a 50-nm insulating layer of Al$_2$O$_3$, and evaporating a metallic top gate (Al, 50 nm thick). While operating the device, we observed the formation of a defect in the Al$_2$O$_3$ layer above the conductive channel next to the gate, as illustrated in Figure~\ref{Fig: Fab}i and seen in the PL image (Figure~\ref{Fig: Fab}j). The stress maps ($\Sigma_Z^{\rm axial}$ in Figure~\ref{Fig: Fab}k and $\sigma_{xz}$ in Figure~\ref{Fig: Fab}l) reveal a compressive stress under the oxide defect ($\Sigma_Z^{\rm axial}\sim20$~MPa) and a tensile stress just outside the defect. The presence of the metallic gate on top of the oxide induces a stress of similar magnitude in the diamond, although the odd parity of $\Sigma_Z^{\rm axial}$ is in contrast with the fully compressive stress observed in Figure~\ref{Fig: Fab}c suggesting a non-trivial mediation by the oxide layer, possibly through permanent deformation of the oxide under the gate while operating the device. These experiments illustrate that the technique can be applied to monitor mechanical deformations in materials and devices outside the diamond, through the elastic stress applied to the diamond substrate as a result, which could find applications as a diagnostic tool for device variability or failure analysis.

\section*{Conclusion} \label{Sec: conclusion}

In summary, we presented a method to quantitatively image the full stress tensor below the surface of a diamond, at room temperature, using a layer of NV strain sensors. We illustrated the versatility of the technique by experimentally imaging the stress under a number of scenarios, from localized implantation damage to nano-indents to devices fabricated on the diamond surface. With our current experimental set-up, the spatial resolution is limited by the diffraction limit ($\approx400$~nm) but could be improved by implementing super-resolution techniques~\cite{Wildanger2011,Chen2017}. The measurement sensitivity is of the order of 1 MPa per $1\times1~\mu$m$^2$ pixel for typical acquisition times ($\sim10$~hours), corresponding to a strain of less than $10^{-6}$~\cite{Barson2017}, and could be improved by an order of magnitude through futher optimisation of the NV-diamond samples~\cite{Kleinsasser2016}. 

The technique could be directly applied to characterise the residual stress in various types of diamond nanostructures commonly used for quantum sensing and quantum information science, such as solid-immersion lenses~\cite{Jamali2014}, nano-pillars~\cite{Babinec2010,Appel2016}, nano-beams~\cite{Khanaliloo2015,Sohn2018}, and photonic waveguides and cavities~\cite{Riedrich-Moller2011,Burek2012,Faraon2012,Hausmann2013}, which could help improve the sensing accuracy or optimise the fabrication processes. Moreover, the technique is compatible with stroboscopic measurements, capable for instance of imaging the time evolution of stress in mechanically driven diamond cantilevers~\cite{Yao2014,Barfuss2015,Ovartchaiyapong2014a}, which are a testbed for hybrid spin-mechanical quantum systems. More generally, the ability to measure the complete stress tensor at the nanoscale opens up the field of multiplexed nanomechanical sensing, where structures with multiple mechanical degrees of freedom could be designed and measured (consider, for example, a cantilever with a paddle on the end), with possible applications in on-chip mass spectroscopy and microfluidics.

\section*{Methods} 

\paragraph*{Diamond Samples}

The NV-diamond samples used in these experiments were made from 4 mm $\times$ 4 mm $\times$ 50 $\mu$m electronic-grade ([N]~$<1$~ppb) single-crystal diamond plates with \{110\} edges and a (001) top facet, purchased from Delaware Diamond Knives. The plates were used as received (i.e. polished with a best surface roughness $<5$~nm Ra) or overgrown with $2~\mu$m of CVD diamond ([N]~$<1$~ppb) using $^{12}$C-enriched (99.95\%) methane~\cite{Teraji2015}, leaving an as-grown surface with a roughness below 1~nm~\cite{Lillie2018}. All the plates were laser cut into smaller 2 mm $\times$ 2 mm $\times$ 50 $\mu$m plates and acid cleaned (15 minutes in a boiling mixture of sulphuric acid and sodium nitrate), and then implanted with either $^{14}$N$^+$ or $^{15}$N$^+$ ions (InnovIon) at an energy of $4-30$ keV and a fluence of $10^{12}-10^{13}$ ions cm$^{-2}$, with a tilt angle of 7$^\circ$. Following implantation, the diamonds were annealed in a vacuum of $10^{-6}-10^{-5}$~Torr to form the NV defects~\cite{Tetienne2018} and acid cleaned (as before). The relevant parameters for the six diamonds used in this work, labelled \#1 to \#6, are listed in the Supporting Information, Table S1.

\paragraph*{Preparation of stress-inducing features}

The scratches examined in Figure~\ref{Fig: intro}g,h and in Figure~\ref{Fig:scratch} were formed by manually dragging the tip of a diamond scribe across the diamond surface (diamond \#1) with moderate pressure~\cite{Lillie2018}. 

The stressed regions induced by implantation in Figure~\ref{Fig: implantation} were produced by evaporating 100 nm of Cr on the diamond surface, patterning the Cr layer via electron-beam lithography and subsequent etching, and implanting the sample with C$_3$ molecules at an energy of 15 keV and a fluence of $3.4\times10^{14}$~molecules cm$^{-2}$ (in diamond \#2). Following implantation the Cr mask was removed with a 2.6:1:20 mixture (by weight) of ceric ammonium nitrate, perchloric acid and H$_2$O. The average penetration depth of the implanted carbon atoms is 10 nm as predicted by Stopping and Range of Ions in Matter (SRIM) simulations, shown in the Supporting Information (Figure S1), with the peak of the vacancy distribution (i.e. of the damage) at about 6 nm depth. In comparison, the average depth of the nitrogen atoms previously implanted to create the NVs was 7 nm in this diamond according to SRIM. The fluence of the carbon implant was such that amorphous carbon was formed (confirmed by Raman spectroscopy) at the peak of the damage, in which NV defects no longer exist, explaining the reduction in PL in the implanted regions (by about 25\%). The remaining PL (75\% of the non-damaged case) is likely to come from NVs that are located under the peak of the damage, i.e. 15-20 nm from the surface, in the tail of the nitrogen implant which extends further than predicted by SRIM due to ion channelling~\cite{Lehtinen2016}. In the Supporting Information (Figure S3-S5), we show additional measurements on a similarly prepared sample except for a higher C$_3$ fluence of $3.4\times10^{15}$~molecules cm$^{-2}$ (in diamond \#6), and including post-implantation annealing in vacuum at 800$^\circ$C for 1 h to induce graphitization of the amorphized implanted regions.   

The indents shown in Figure~\ref{Fig:indent} were formed by applying a load of 1 N using a Berkovich tip (three-sided pyramid), corresponding to a maximum vertical displacement of about 130 nm, and a residual displacement of about 15 nm (plastic deformation). Loads of 500 mN (70 nm maximum displacement) and under did not plastically deformed the diamond. A series of 10 indents were made with a 1 N load for each and a lateral spacing of $20~\mu$m, showing good reproducibility in the load-unload curve. 

The devices imaged in Figure~\ref{Fig: Fab} were made as follows. The metallic wire in Figure~\ref{Fig: Fab}a-d (diamond \#4) was made via photolithography, evaporation of a Ti/Au stack (thickness 10/100 nm) and lift-off. The embedded wire in Figure~\ref{Fig: Fab}e-h (diamond \#5) was made by patterning an evaporated stack of Ti/Pt/Au (thickness 10/10/70\,nm) and annealing the sample at 600$^\circ$C for 20 minutes in hydrogen gas (10 Torr) to favor inter-diffusion of Ti and C atoms, which formed a TiC layer extending about 15 nm into the diamond~\cite{Broadway2018}. The transistor devices in Figure~\ref{Fig: Fab}i-l were made on the same diamond (diamond \#5) by making pairs of TiC/Pt/Au contacts (source and drain), hydrogen-terminating the diamond surface via plasma treatments to form a conductive channel between two TiC/Pt/Au contacts~\cite{Broadway2018}, depositing a 50-nm layer of Al$_2$O$_3$ on the whole sample via atomic layer deposition (ALD), and finally patterning an evaporated layer of Al (thickness 50 nm) to form the top gate. 

\paragraph*{Experimental apparatus}

The diamonds were placed on a glass cover slip patterned with a metallic waveguide for microwave (MW) delivery and connected to a printed circuit board (PCB). A layer of immersion oil was used to ensure good optical transmission between the diamond and the glass cover slip, with the NV side of the diamond exposed to ambient air. 

The NV measurements were performed using a custom-built wide-field fluorescence microscope~\cite{Simpson2016,Broadway2018}. Optical excitation from a $\lambda = 532$~nm continuous-wave (CW) laser (Coherent Verdi) was gated using an acousto-optic modulator (AA Opto-Electronic MQ180-A0,25-VIS), beam expanded (5x) and focused using a wide-field lens ($f = 200$~mm) to the back aperture of an oil immersion objective lens (Nikon CFI S Fluor 40x, NA = 1.3). The CW laser power entering the objective was 300 mW, corresponding to a maximum power density of about 5~kW cm$^{-2}$ at the sample given the $\approx120~\mu$m~$1/e^2$ beam diameter. The photoluminescence (PL) from the NV defects was separated from the excitation light with a dichroic mirror and filtered using a bandpass filter before being imaged using a tube lens ($f = 300$ mm) onto an sCMOS camera (Andor Zyla 5.5-W USB3). MW excitation was provided by a signal generator (Rohde \& Schwarz SMBV100A) gated using the built-in IQ modulation and amplified (Mini-Circuits HPA-50W-63) before being sent to the PCB. A pulse pattern generator (SpinCore PulseBlasterESR-PRO 500 MHz) was used to gate the excitation laser and MW and to synchronise the image acquisition.

The optically-detected magnetic resonance (ODMR) spectra of the NV layer were obtained by sweeping the microwave frequency while repeating the following sequence: $10~\mu$s laser pulse, $1.5~\mu$s wait time, $300$~ns microwave pulse; and alternating MW on/off to remove the common-mode PL fluctuations, with total acquisition times of at least 10 hours typically. All measurements were performed in an ambient environment at room temperature, under a bias magnetic field $\vec{B}$ generated using a permanent magnet.

\section*{Acknowledgements}

We acknowledge support from the Australian Research Council (ARC) through grants DE170100129, CE170100012, FL130100119 and DP170102735. This work was performed in part at the Materials Characterisation and Fabrication Platform (MCFP) at the University of	Melbourne and the Victorian Node of the Australian National Fabrication Facility (ANFF). This work was performed in part at the Melbourne Centre for Nanofabrication (MCN) in the Victorian Node of the Australian National Fabrication Facility (ANFF). We acknowledge the AFAiiR node of the NCRIS Heavy Ion Capability for access to ion-implantation facilities. D.A.B., S.E.L., D.J.M. and A.T. are supported by an Australian Government Research Training Program Scholarship. T.T. acknowledges the support of Grants-in-Aid for Scientific Research (Grant Nos. 15H03980, 26220903, and 16H06326), the ``Nanotechnology Platform Project'' of MEXT, Japan, and CREST (Grant No. JPMJCR1773) of JST, Japan. \\


\clearpage

\begin{widetext}

\section*{Supporting Information}

\section{Analysis of the ODMR data} \label{Sec: ODMR analysis}

The ODMR data is analyzed by first fitting the ODMR spectrum at each pixel with a sum of eight Lorentzian functions with free frequencies, line widths and amplitudes. The resulting frequencies $\{\omega_{\pm,i}\}$ (with $i=1\dots4$) are then used to infer the Hamiltonian parameters (denoted as vector $\vec{p}$) by minimising the root-mean-square error function
\begin{equation} \label{Eq:error}
\varepsilon(\vec{p}) = \sqrt{\frac{1}{8}\sum_{i=1}^4 \sum_{j=\pm} \left[\omega_{j,i} - \omega_{j,i}^{\rm calc}(\vec{p}) \right]^2}
\end{equation}
where $\{\omega_{\pm,i}^{\rm calc}(\vec{p})\}$ are the calculated frequencies for a given set of parameters $\vec{p}$ obtained by numerically computing the eigenvalues of the spin Hamiltonian for each NV orientation,
\begin{eqnarray} \label{eq:Ham2}
\begin{aligned}
H_i &= (D + \mathcal{M}_{Z_i}+k_\parallel E_{Z_i})S_{Z_i}^2   \\
& \quad + \gamma_{\rm NV}(B_{X_i}S_{X_i}+B_{Y_i}S_{Y_i}+B_{Z_i}S_{Z_i})  \\
& \quad -\left(\mathcal{M}_{X_i}+k_\perp E_{X_i}\right) ( S_{X_i}^2 - S_{Y_i}^2) \\
& \quad + \left(\mathcal{M}_{Y_i}+k_\perp E_{Y_i}\right)( S_{X_i}S_{Y_i} + S_{Y_i}S_{X_i}), 
\end{aligned}
\end{eqnarray} 
where for completeness we added the effect of the true electric field, $\vec{E}=(E_{X_i},E_{Y_i},E_{Z_i})$ expressed in the NV frame ($X_iY_iZ_i$ coordinate system), with electric susceptibility parameters $k_\parallel = 0.35(2)$ Hz~cm~V$^{-1}$ and $k_\perp= 17(3)$ Hz~cm~V$^{-1}$~\cite{Dolde2011,Doherty2012}. We note that the additional strain terms derived in Ref.~\cite{Udvarhelyi2018} were neglected as they are an order of magnitude smaller.

In general, the set of four Hamiltonians $\{H_1,H_2,H_3,H_4\}$ is characterized by a total of 13 parameters: the temperature-dependent zero-field splitting parameter $D$, the magnetic field $\vec{B}=(B_x,B_y,B_z)$, the electric field $\vec{E}=(E_x,E_y,E_z)$, and the stress tensor $\overleftrightarrow\sigma=\{\sigma_{xx},\sigma_{yy},\sigma_{zz},\sigma_{yz},\sigma_{xz},\sigma_{xy}\}$ (in Voight notation), all expressed in the crystal frame ($xyz$ coordinate system). The relationships between these parameters and the quantities entering the Hamiltonians are given by Equation~(2-4) of the main text for the effect of stress, and by
\begin{eqnarray} \label{eq:B}
B_{X_i} &=& \frac{1}{\sqrt{6}}(-f_ig_iB_x-g_iB_y-2f_iB_z) \nonumber \\
B_{Y_i} &=& \frac{1}{\sqrt{2}}(f_ig_iB_x-g_iB_y) \\
B_{Z_i} &=& \frac{1}{\sqrt{3}}(f_ig_iB_x+g_iB_y-f_iB_z) \nonumber
\end{eqnarray}
for the magnetic field (which simply expresses the rotation of the coordinate system), with identical expressions for the electric field.

To allow inference of these 13 parameters from the 8 ODMR frequencies, we must make assumptions, which may introduce systematic errors in the inferred stress values (see Section~\ref{Sec: accu}). First, we assume that there is no net electric field, i.e. $\vec{E}=0$. In reality, an electric field may be present in the NV layer as a result of charge effects in the diamond bulk and band bending at the diamond surface~\cite{Broadway2018,Mittiga2018}. We emphasize that the effect of stress and electric field cannot be combined in a single entity when measuring multiple NV orientations, even though they both effectively act as an electric field in the Hamiltonian, Equation~(\ref{eq:Ham2}). This is because the effective electric field associated with stress differs for different NV orientations since it derives from a tensor, see Equation~(2) of the main text, while the true electric field is a vector that is simply projected onto the NV axes, see Equation~(\ref{eq:B}). For the same reason, an extra four NV orientations (inverting the nitrogen and the vacancy) would need to be considered in the presence of a true electric field, whereas the effective electric field associated with stress, $\vec{\cal M}_i$, is invariant under this inversion.  

Second, we assume that the stress is null except near the deliberately introduced features (scratches, implantation damage, indents etc.). In reality, there may exist a residual stress across the sample~\cite{Bauch2018} (see Section~\ref{Sec: accu}). This assumption allows us to use the ODMR data far from those features to estimate $D$ and $\vec{B}$, by minimising $\varepsilon(D,\vec{B})$. Doing so to the full field of view (excluding the stressed regions) gives a value of $D$ that is relatively uniform across the image (typically between 2869.5 and 2870.5 MHz), whereas $\vec{B}$ exhibits slow variations due to the magnetic field produced by the permanent magnet being non uniform. A multi-polynomial fit was applied to the $\vec{B}$ maps, and this fit was then used as a known input for the second fitting step. In the second step, we refit the data (now including the stressed regions) with the six stress parameters as the only unknown, i.e. we minimise $\varepsilon(\overleftrightarrow\sigma)$, while holding $D$ and $\vec{B}$ to the values determined previously.

\section{Uniqueness of the solution}\label{Sec: unique}

When minimising $\varepsilon(\overleftrightarrow\sigma)$ to infer the best-fit stress tensor $\overleftrightarrow\sigma$ at each pixel, we rely on the fact that there is a unique solution to the optimisation problem. Here we investigate the conditions in which this is true. We first consider the small magnetic field limit ($|\vec{B}|\ll D$) leading to Equation~(5) of the main text, allowing us to obtain simple expressions linking the NV frequencies $(\omega_\pm)_i$ to the stress components. Namely, for the sum frequencies $S_i=(\omega_++\omega_-)_i$ we have
\begin{eqnarray} \label{eq:omega2}
\begin{aligned}
S_i &= D+a_1(\sigma_{xx}+\sigma_{yy}+\sigma_{zz}) \\
 & \quad +2a_2(f_i\sigma_{xy} -g_i \sigma_{xz} -f_ig_i \sigma_{yz}) 
 \end{aligned}
\end{eqnarray} 
which gives a set of four independent linear equations with four unknowns: $\sigma_{xy}$, $\sigma_{xz}$, $\sigma_{yz}$ and the sum $D+a_1\Sigma_Z^{\rm axial}$ where $\Sigma_Z^{\rm axial}=\sigma_{xx}+\sigma_{yy}+\sigma_{zz}$. Therefore, these four quantities can always be uniquely retrieved. 

On the other hand, the difference frequencies $D_i=(\omega_+-\omega_-)_i$ are given by 
\begin{eqnarray} \label{eq:omega3}
\begin{aligned}
D_i^2 &= \frac{4\gamma_{\rm NV}^2}{3}(f_ig_iB_x+g_iB_y-f_iB_z)^2 \\
& \quad +4[b(-\sigma_{xx}-\sigma_{yy}+2\sigma_{zz})  \\
& \quad ~~~~~ + c(2f_i \sigma_{xy}+g_i\sigma_{xz}+f_ig_i\sigma_{yz})]^2  \\
& \quad +12[b(\sigma_{xx} - \sigma_{yy}) + c(g_i\sigma_{xz}-f_ig_i \sigma_{yz})]^2.  
\end{aligned}
\end{eqnarray} 
This gives a set of four equations with only two unknown, for instance $\sigma_{xx}$ and $\sigma_{yy}$, where $\sigma_{zz}$ can be expressed as a function of $(\sigma_{xx},\sigma_{yy})$ and the measured $\{S_i\}$, whereas the shear stress components have been completely determined from $\{S_i\}$. However, these four equations are not linear and not necessarily independent with respect to the unknown parameters. In particular, in the situation where all the shear components are null, $\sigma_{xy}=\sigma_{xz}=\sigma_{yz}=0$, Equation~(\ref{eq:omega3}) reduces to 
\begin{eqnarray} \label{eq:omega4}
\begin{aligned}
D_i^2 &= \frac{4\gamma_{\rm NV}^2}{3}(f_ig_iB_x+g_iB_y-f_iB_z)^2 \\
& \quad +4b^2(-\sigma_{xx}-\sigma_{yy}+2\sigma_{zz})^2  \\
& \quad +12b^2(\sigma_{xx} - \sigma_{yy})^2.  
\end{aligned}
\end{eqnarray} 
Clearly, in this situation only a single combination of $\sigma_{xx}$ and $\sigma_{yy}$ can be inferred from the measured $\{D_i\}$, which means that these axial stress components cannot be resolved individually, i.e. there an infinite number of solutions for the axial stress components. 

When the shear components are non-zero, however, Equation~(\ref{eq:omega3}) forms an overdetermined system that is likely to have a unique solution. To test this, we numerically explored the full parameter space for the six stress components and evaluated the error $\varepsilon$ comparing the NV frequencies calculated with a given ${\overleftrightarrow\sigma}_0$ versus those calculated with a trial solution ${\overleftrightarrow\sigma}_{\rm trial}$, while holding the other parameters $D$ and $\vec{B}$ constant. We found that there is indeed a unique solution that gives a vanishing error, ${\overleftrightarrow\sigma}_{\rm trial}={\overleftrightarrow\sigma}_0$, except when two of the three axial shear components are null (in which case there are two solutions, where the incorrect solution simply swaps two of the axial components) and of course when all three shear components are null (as discussed above). This conclusion holds regardless of the value of the magnetic field or of the other stress components. Thus, our method enables the full stress tensor to be determined with no ambiguity, as long as at least two of the shear stress components are not exactly null.  

\section{Sensitivity of the method}\label{Sec: sens}

The sensitivity of the stress measurements, which dictates the smallest detectable change (in space or time) for a given integration time, is limited by statistical errors arising from the statistical noise in the measured NV frequencies $(\omega_\pm)_i$. The latter originates from photon count noise in the ODMR data, which has contributions from both the photon shot noise and dark current noise from the camera. To characterise this noise, we calculate the standard deviation from an ensemble of pixels in a region of the image where the stress is relatively uniform, i.e. the pixel-to-pixel noise. We find this noise to be typically of the order of 1 MPa for the shear components and 10 MPa for the axial components for a $1\times1~\mu$m$^2$ pixel and a total integration time of 10 hours, with no significant improvement with longer acquisition indicating that this is close to the technical noise floor. As expected, the error is larger on the axial stress components than on the shear components, because of the difference in sensitivity of the NV frequencies to these different components. Indeed, the shear components are obtained from Equation~(\ref{eq:omega3}) as a linear combination of the sum frequencies $\{S_i\}$, with a sensitivity
\begin{eqnarray} \label{eq:sensitivity}
\frac{\partial S_i}{\partial \sigma_{xy}} \sim a_2 \approx 4~{\rm MHz~GPa^{-1}}.
\end{eqnarray}
In contrast, the individual axial components are derived from the difference frequencies $\{D_i\}$ which has a sensitivity, in the limit $\gamma_{\rm NV}B_{Z_i}\gg{\cal M}_{\perp,i}$ \cite{Bauch2018},
\begin{eqnarray} \label{eq:sensitivity2}
\frac{\partial D_i}{\partial \sigma_{xx}} \sim b\frac{{\cal M}_\perp}{\gamma_{\rm NV}B_{Z_i}} \approx 2~{\rm MHz~GPa^{-1}}\times \frac{{\cal M}_\perp}{\gamma_{\rm NV}B_{Z_i}},
\end{eqnarray}
where ${\cal M}_\perp\doteq \sqrt{{\cal M}_{X_i}^2+{\cal M}_{Y_i}^2} $. The suppression factor $\frac{{\cal M}_\perp}{\gamma_{\rm NV}B_{Z_i}}$ depends on the exact conditions but is typically below 0.1, which results in a statistical noise on the axial components (as obtained by solving the overdetermined problem) typically an order of magnitude larger than that on the shear components. We note that the sum of the axial components, $\Sigma_Z^{\rm axial}=\sigma_{xx}+\sigma_{yy}+\sigma_{zz}$, is fixed by the sum frequencies $\{S_i\}$ and therefore is not affected by this suppression factor, yielding a statistical noise down to 1 MPa. This was employed in Figure 5 of the main text, where we plotted $\Sigma_Z^{\rm axial}$ in order to detect weak stress features. For applications where measuring the individual axial stress components with maximum sensitivity is desirable, we could perform four consecutive ODMR measurements with an external magnetic field optimized to minimise $B_{Z_i}$ for each NV orientation sequentially. Such a sequential measurement strategy would also allow much larger stress values to be measured (up to 100 GPa), currently limited to a few GPa by the minimum spacing between adjacent ODMR lines (20-30 MHz)~\cite{Tetienne2018b}.

\section{Accuracy of the method}\label{Sec: accu}

Systematic errors in the stress values may be present as a result of the assumptions made in the model converting the frequencies $(\omega_\pm)_i$ into stress. 
One assumption is the neglect of the true electric field, $\vec{E}$. In Ref.~\cite{Broadway2018} it was reported that in samples comparable to those used here, an electric field perpendicular to the diamond surface exists owing to surface band bending, with values up to $E_z\sim500$~kV cm$^{-1}$. We first examine the effect of a non-zero $E_z$ on the determination of the shear stress components. Extending Equation~(\ref{eq:omega2}) to include $E_z$ gives  
\begin{eqnarray} \label{eq:shearEz}
\begin{aligned}
S_i &= D+a_1\Sigma_Z^{\rm axial}-\frac{k_\parallel}{\sqrt{3}}f_iE_z \\
 & \quad +2a_2(f_i\sigma_{xy} -g_i \sigma_{xz} -f_ig_i \sigma_{yz}).
 \end{aligned}
\end{eqnarray}
We can then write the following linear combinations
\begin{eqnarray} \label{eq:shearEz2}
\begin{aligned}
S_1-S_2-S_3+S_4 &= 8a_2\sigma_{xy}-\frac{4k_\parallel}{\sqrt{3}}E_z \\
S_1-S_2+S_3-S_4 &= 8a_2\sigma_{xz} \\
S_1+S_2-S_3-S_4 &= 8a_2\sigma_{yz} \\
S_1+S_2+S_3+S_4 &= 4D+4a_1\Sigma_Z^{\rm axial}.
 \end{aligned}
\end{eqnarray}
Thus, the presence of $E_z$ will affect only the $\sigma_{xy}$ component. Precisely, the assumption $E_z=0$ in the stress reconstruction amounts to offsetting $\sigma_{xy}$ by $-\frac{k_\parallel}{2a_2\sqrt{3}}E_z$ where $a_2<0$. For instance, the $\approx10$~MPa background observed in the $\sigma_{xy}$ map in Figure~2d of the main text can be attributed to an electric field of $E_z\approx+350$~kV cm$^{-1}$ uniform across the field of view, in line with the values reported in Ref.~\cite{Broadway2018}. Furthermore, since the magnetic field does not enter Equation~(\ref{eq:shearEz2}), the assumption that there is no background stress away from the local features should not affect the outcome for these four quantities, which then correspond to the net stress including a non-zero background contribution if present. 

This is in contrast with the individual axial components, which are determined through Equation~(\ref{eq:omega3}) which involves in a non-trivial way both the electric field and the magnetic field, which itself is determined under the assumption of no background electric field or stress. For illustration purpose, let us consider the situation where $\sigma_{xx}$ is the only non-zero stress component. Equation~(\ref{eq:omega3}) then reads
\begin{eqnarray} \label{eq:axialEz}
\begin{aligned}
D_i^2 &= 4(\gamma_{\rm NV} B_{Z_i})^2 +4(-b\sigma_{xx}-\frac{2f_i}{\sqrt{6}}k_\perp E_z)^2  \\
& \quad +4(\sqrt{3}b\sigma_{xx})^2 
\end{aligned}
\end{eqnarray} 
Here, an error on $B_{Z_i}$ would act as a simple offset on $\sigma_{xx}$, while $E_z$ has a more subtle effect as it can add or subtract to the $\sigma_{xx}$ term depending on the NV orientation. Moreover, when the other stress components are present, an error on the magnetic field may also act to bias the estimated stress values in a way that depends on the other stress values. In other words, the assumptions made may introduce not just a global offset on the stress maps but also a distortion of the spatial features. 

To test these effects, we used a set of experimental data and compared the stress maps produced by different analysis procedures. Namely, instead of assuming no electric field and background stress, we estimated the magnetic field (far from the local stress features) by including an electric field along $z$, i.e. minimising $\varepsilon(D,\vec{B},E_z)$, or by including a background axial stress along $z$, i.e. minimising $\varepsilon(D,\vec{B},\sigma_{zz})$, or by including both at the same time. The stress tensor was then determining by fitting the whole image again (but including the stressed regions) while holding $D$, $\vec{B}$ and $E_z$ to the previously determined values. As expected, the shear components as well as $\Sigma_Z^{\rm axial}$ are essentially unchanged by the analysis procedure, except for a global offset in $\sigma_{xy}$ due to the electric field. On the other hand, these various methods affect the axial stress components not only as a global offset, but also by changing the values near local features. For example, in Figure 2e of the main text, the increase in $\sigma_{xx}$ near the implanted region can vary from 20 MPa to nearly zero depending on the initial assumptions, compensated by an opposite variation in $\sigma_{yy}$ or $\sigma_{zz}$ (since the sum is well determined). This means that the uncertainty on these individual axial components can be up to 100\%, calling for caution when extracting quantitative information from the corresponding maps. Nevertheless, the fact that the stress maps in Figure 2e of the main text exhibits the expected symmetry (e.g., $\sigma_{xx}$ maximum on the $x$ axis and minimum on the $y$ axis) indicates that the different axial components have been sensibly separated. 

Finally, we note that the uncertainty on the spin-mechanical coupling parameters adds an additional uncertainty on the stress values. Namely, the 5\% uncertainty on $a_2$ translates into a similar uncertainty on the shear stress components, whereas the axial components are affected by the $>10\%$ uncertainty on $b$.

\section{Body force approximations} \label{Sec: Body force approximations}

The body force equations are given by
\begin{align}
- f_x &= \partial_x \sigma_{xx} + \partial_y \sigma_{xy} + \partial_z \sigma_{xz} \nonumber \\
- f_y &= \partial_x \sigma_{xy} + \partial_y \sigma_{yy} + \partial_z \sigma_{yz} \\
- f_z &= \partial_x \sigma_{xz} + \partial_y \sigma_{yz} + \partial_z \sigma_{zz}. \nonumber
\end{align}
The measured stress components are averaged over the thickness of the NV layer
\begin{equation}
\bar{\sigma}_{ij} (x, y) = \frac{1}{\Delta} \int_{z_0 - \Delta/2}^{z_0 - \Delta/2} \sigma_{ij} dz
\end{equation}
where $z_0$ and $\Delta$ are the mean depth and thickness of the NV layer, assuming that the NV defects are uniformly distributed within the slab. Our aim is to use the stress equations to express the averaged body forces
\begin{equation}
\bar{f}_{i} (x, y) = \frac{1}{\Delta} \int_{z_0 - \Delta/2}^{z_0 - \Delta/2} f_{i} dz
\end{equation}
in terms of the measured averaged stress components $\bar{\sigma}_{ij}$ so that we can use the measurements to obtain the average body forces and interpret their origin.

To do this, let's integrate the stress equations

\begin{equation}
\begin{aligned}
-\frac{1}{\Delta }\int _{z_0 - \Delta/2}^{z_0 + \Delta /2} f_x dz &=
\frac{1}{\Delta } \int _{z_0-\Delta /2}^{z_0 + \Delta/2} \frac{\partial \sigma _{xx}}{\partial x} + \frac{\partial \sigma _{xy}}{\partial y} + \frac{\partial \sigma _{xz}}{\partial z} dz   \\
-\frac{1}{\Delta }\int _{z_0 - \Delta/2}^{z_0 + \Delta /2} f_y dz &=
\frac{1}{\Delta } \int _{z_0-\Delta /2}^{z_0 + \Delta/2} \frac{\partial \sigma _{xy}}{\partial x} + \frac{\partial \sigma _{yy}}{\partial y} + \frac{\partial \sigma _{yz}}{\partial z} dz   \\
-\frac{1}{\Delta }\int _{z_0 - \Delta/2}^{z_0 + \Delta /2} f_z dz &=
\frac{1}{\Delta } \int _{z_0-\Delta /2}^{z_0 + \Delta/2} \frac{\partial \sigma _{xz}}{\partial x} + \frac{\partial \sigma _{yz}}{\partial y} + \frac{\partial \sigma _{zz}}{\partial z} dz   \nonumber
\end{aligned}
\end{equation}
which gives
\begin{equation}
\begin{aligned}
- \bar{f}_x &= \partial_x \bar{\sigma}_{xx} + \partial_y \bar{\sigma}_{xy} + \frac{1}{\Delta} [\sigma_{xz}(z_0 + \frac{\Delta}{2})  -  \sigma_{xz}(z_0 - \frac{\Delta}{2})]\\
- \bar{f}_y &= \partial_x \bar{\sigma}_{xy} + \partial_y \bar{\sigma}_{yy} +\frac{1}{\Delta} [\sigma_{yz}(z_0 + \frac{\Delta}{2})  -  \sigma_{yz}(z_0 - \frac{\Delta}{2})] \\
- \bar{f}_z &= \partial_x \bar{\sigma}_{xz} + \partial_y \bar{\sigma}_{yz} +\frac{1}{\Delta} [\sigma_{zz}(z_0 + \frac{\Delta}{2})  -  \sigma_{zz}(z_0 - \frac{\Delta}{2})]. \nonumber
\end{aligned}
\end{equation}
If $\Delta$ is sufficiently small (i.e. NV layer sufficiently thin), then we could immediately apply the above by making the approximation that each third term of the right hand side is negligible,
\begin{equation}\label{Eq: force approx}
\begin{aligned}
- \bar{f}_x &= \partial_x \bar{\sigma}_{xx} + \partial_y \bar{\sigma}_{xy}\\
- \bar{f}_y &= \partial_x \bar{\sigma}_{xy} + \partial_y \bar{\sigma}_{yy} \\
- \bar{f}_z &= \partial_x \bar{\sigma}_{xz} + \partial_y \bar{\sigma}_{yz}.
\end{aligned}
\end{equation}
This is equivalent to stating that the out-of-plane ($z$) derivatives of the respective stress components are negligible at $z_0$.
Alternatively, we can appeal to Leibniz integral rule to identify
\begin{equation}
\frac{d}{dz_0}\bar{\sigma}_{ij}=\frac{1}{\Delta }\frac{d}{dz_0}\int _{z_0 - \Delta/2}^{z_0 + \Delta /2}\sigma _{ij}dz =\frac{1}{\Delta }(\sigma _{ij} (z_0+\Delta)-\sigma _{ij}(z_0)). \nonumber
\end{equation}
Thus, the neglect of the third term in the above may instead be interpreted as stating that the average stress components depend little upon the depth of the NV layer $z_0$ (i.e. $\Delta$ is sufficiently large to capture all $z$ dependence).
So, in either limit -- $\Delta$ much smaller than $z$ variations or $\Delta$ much larger than $z$ variations -- we can adopt the approximate equation, Equation~(\ref{Eq: force approx}).

In intermediate situations, one can derive an estimate for the $z$ derivatives as follows. If we assume that the stress decays in the $z$ direction as $\sigma_{ij}(z)=\sigma_{ij}(0)\exp(-z/L)$ where $L$ is the characteristic decay length, then for $L\gg z_0,\Delta$ we can approximate the $z$ derivative term as
\begin{eqnarray}
\frac{1}{\Delta} [\sigma_{ij}(z_0 + \frac{\Delta}{2})  -  \sigma_{ij}(z_0 - \frac{\Delta}{2})] &=& \frac{\sigma_{ij}(z_0)}{\Delta}[e^{- \frac{\Delta}{2L}}-e^{+ \frac{\Delta}{2L}}] \\
&\approx & -\frac{\bar{\sigma}_{ij}}{L}
\end{eqnarray}
hence 
\begin{equation}\label{Eq: force approx2}
\begin{aligned}
- \bar{f}_x &= \partial_x \bar{\sigma}_{xx} + \partial_y \bar{\sigma}_{xy} -\frac{\bar{\sigma}_{xz}}{L}\\
- \bar{f}_y &= \partial_x \bar{\sigma}_{xy} + \partial_y \bar{\sigma}_{yy} -\frac{\bar{\sigma}_{yz}}{L} \\
- \bar{f}_z &= \partial_x \bar{\sigma}_{xz} + \partial_y \bar{\sigma}_{yz} -\frac{\bar{\sigma}_{zz}}{L}.
\end{aligned}
\end{equation}

$L$ is expected to be similar to the characteristic decay length observed in the $xy$ plane, i.e. a few microns. In practice, we find that the third term in the right hand side of Equation~(\ref{Eq: force approx2}) is negligible in calculating $f_x$ and $f_y$, but is dominant in calculating $f_z$, i.e. we have $\bar{f}_z \approx \frac{\bar{\sigma}_{zz}}{L}$. In Figure 3 and 4 of the main text, the body force were calculated using Equation~(\ref{Eq: force approx2}), taking $L=3~\mu$m.

\section{Diamond samples} 

The relevant parameters for the six diamonds used in this work, labelled \#1 to \#6, are listed in Table \ref{Tab: diamond samples}, indicating the type of surface (polished or as-grown) as well as the parameters of the nitrogen implant used to create the NV defects.

\begin{table}[h!]
	\begin{tabular}{c | c | c | c | c }
		Diamond & Surface & Energy & Fluence       & Figure \\
				&         & (keV)  & (ions/cm$^2$) &         \\
		\hline
		\#1 & A & 6 & $10^{13}$ & 1, 4 \& S7 \\ 
		\#2 & A & 4 & $10^{13}$  & 2, S2 \& S3 \\ 
		\#3 & A & 4 & $10^{13}$  & 3 \& S6 \\ 
		\#4 & P & 4 &  $10^{13}$  & 4a-d \\ 
		\#5 & P & 30 & $3\times10^{12}$ & 4e-l  \\ 
		\#6 & A & 4 & $10^{13}$  & S3, S4 \& S5 \\   
	\end{tabular}
\caption{List of diamonds used in this work, indicating the type of surface (polished `P', or as-grown `A'), the implantation energy and fluence of the nitrogen implant used to create the NV defects, and the figure in which each sample appears in the main text and in this document. Samples \#2 and \#6 were also patterned and implanted with 5 keV C to a fluence of $10^{15}$ and $10^{16}$~carbon cm$^{-2}$, respectively.}
\label{Tab: diamond samples}
\end{table}

\section{Supplementary data: Implantation damage} \label{Sec: Implantation charac}

\subsection{SRIM simulations} \label{Sec: Different implantation shapes}

We performed full cascade Stopping and Range of Ions in Matter (SRIM) Monte Carlo simulations to estimate the depth distribution of the damage created by C implantation with respect to the NV layer, assuming a diamond density of 3.51 g~cm$^{-3}$ and a displacement energy of 50~eV. Figure~\ref{Fig: SRIM}a shows the depth distribution of the N atoms implanted at 4 keV as in diamond \#2, which peaks at 7 nm. A C$_3$ molecule implant at 15 keV (fluence $3.4\times10^{14}$~molecules~cm$^{-2}$) was performed to achieve a 5 keV C implant to a fluence of $10^{15}$~atoms~cm$^{-2}$. Figure~\ref{Fig: SRIM}b shows the vacancy and C depth distributions from this implant which peak at about 6 and 10 nm, respectively. Due to the large vacancy concentration, amorphous carbon forms at the peak of the damage (the critical vacancy concentration for amorphisation in diamond is $2\times10^{22}$~vacancies~cm$^{-3}$~\cite{Fairchild2012}), as confirmed by Raman spectroscopy (not shown). We note that although SRIM is able to model the general shape of the implant profile well, it does not account for channeling effects and may therefore underestimate the ion range~\cite{Lehtinen2016}. 

\begin{figure}[h!]
	\includegraphics[width=0.4\textwidth]{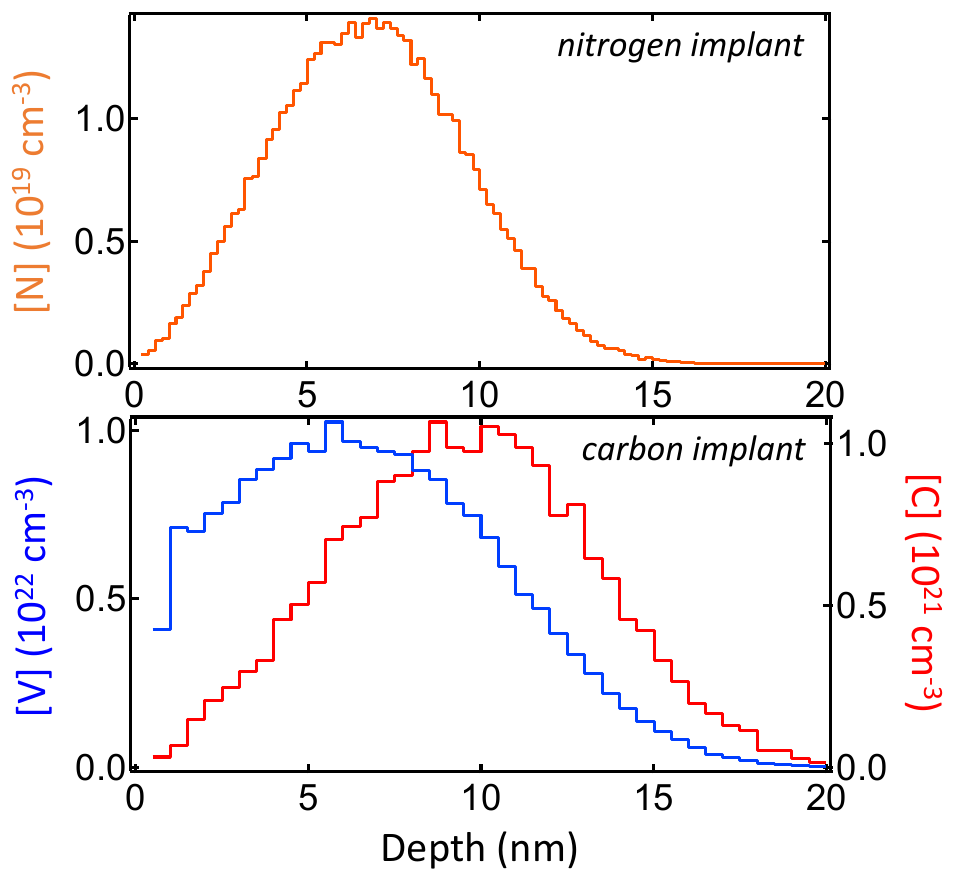}
	\caption{Stopping and Range of Ions in Matter (SRIM) simulations of the N implant used to create the NV defects (a) and of the C implant used to create the damage (b). Plotted are the nitrogen concentration (a), the vacancy concentration (b, blue data) and the implanted carbon concentration (b, red data), as a function of depth. 
	}
	\label{Fig: SRIM}
\end{figure}

\subsection{Stress from different carbon implant shapes} \label{Sec: Different implantation shapes}

Various structures were fabricated with the C$_3$ implant in diamond \#2, imaged in Figure~\ref{Fig: implantation shapes} showing the PL image, a stress map ($\sigma_{xy}$, which has the largest signal-to-noise ratio) and a line cut for different shapes. The square ring (Figure~\ref{Fig: implantation shapes}a,b) has a very similar stress pattern to the circular ring and has a maximum shear stress magnitude of $|\sigma_{xy}|\approx 15$~MPa localised at the edges of the damaged region. Figure~\ref{Fig: implantation shapes}c,d show the case of closed squares of different sizes, indicating that the maxima coincide with the edges of the carbon implant with a small increase of the magnitude of shear stress with increasing size. The smallest squares ($1\times 1~\mu$m) are still resolvable on the right of the image. Lines of different widths are imaged in Figure~\ref{Fig: implantation shapes}e,f, showing more clearly the size dependence. A $1~\mu$m-wide line following a zigzag pattern is imaged in Figure~\ref{Fig: implantation shapes}g,h, clearly showing that lines separated by less than $1~\mu$m can be resolved, allowing us to conclude that the spatial resolution of the technique is at least better than $1~\mu$m.  

\begin{figure}[htb!]
	\includegraphics[width=0.9\textwidth]{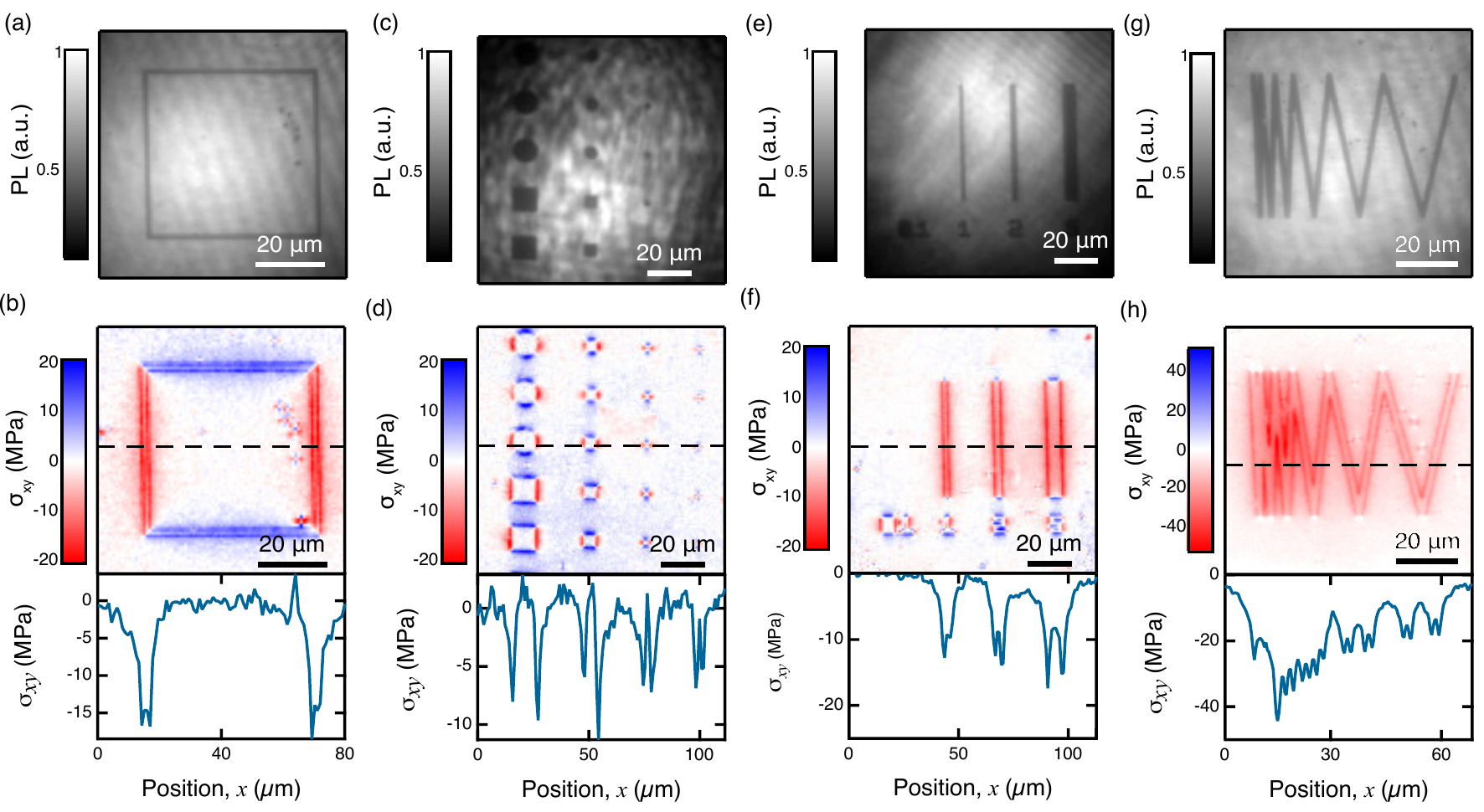}
	\caption{Stress induced from different shapes of carbon implant in diamond \#2.
		(a) NV PL of a square ring, with a wall width of 1~$\mu$m.
		(b) Measured stress component, $\sigma_{xy}$, from the square ring shown in (a) with a line cut along the dashed line.
		(c) NV PL of a series of solid squares and circles varying in width/radius from 10~$\mu$m (left) to 1~$\mu$m (right). 
		(d)  $\sigma_{xy}$ stress from (c) with a line cut along the dashed line. 
		(e) NV PL of a series of solid lines varying in width from 0.1~$\mu$m (left) to 5~$\mu$m (right).
		(f)   $\sigma_{xy}$ stress from (e) with a line cut along the dashed line. 
		(g) NV PL of a solid lines with a width of 1~$\mu$m in a zigzag pattern. 
		(h)  $\sigma_{xy}$ stress from (g) with a line cut along the dashed line. 
	}
	\label{Fig: implantation shapes}
\end{figure}

\subsection{Effect of carbon implant fluence and post-annealing}\label{Sec: Annealing}

We fabricated similar structures in a separate diamond (diamond \#6) nominally identical to the previous one (in particular same parameters for the nitrogen implant used to create the NVs), but increasing the fluence of the C$_3$ implant to $3.4\times10^{15}$~molecules/cm$^2$. Despite this fluence being ten times as large as previously, we found the measured stress to be actually smaller than in the low fluence sample, by roughly a factor of 2, as shown in Figure~\ref{Fig: dose}. This can be explained by a difference in stress relief via the free surface, which was confirmed by AFM (Figure~\ref{Fig: implant charactrisation}a) where the bulge of the diamond surface in diamond~\#6 was 10 nm high (Figure~\ref{Fig: implant charactrisation}b) against 2 nm in diamond~\#2. 

\begin{figure}[t!]
	\centering
	\includegraphics[width=0.9\columnwidth]{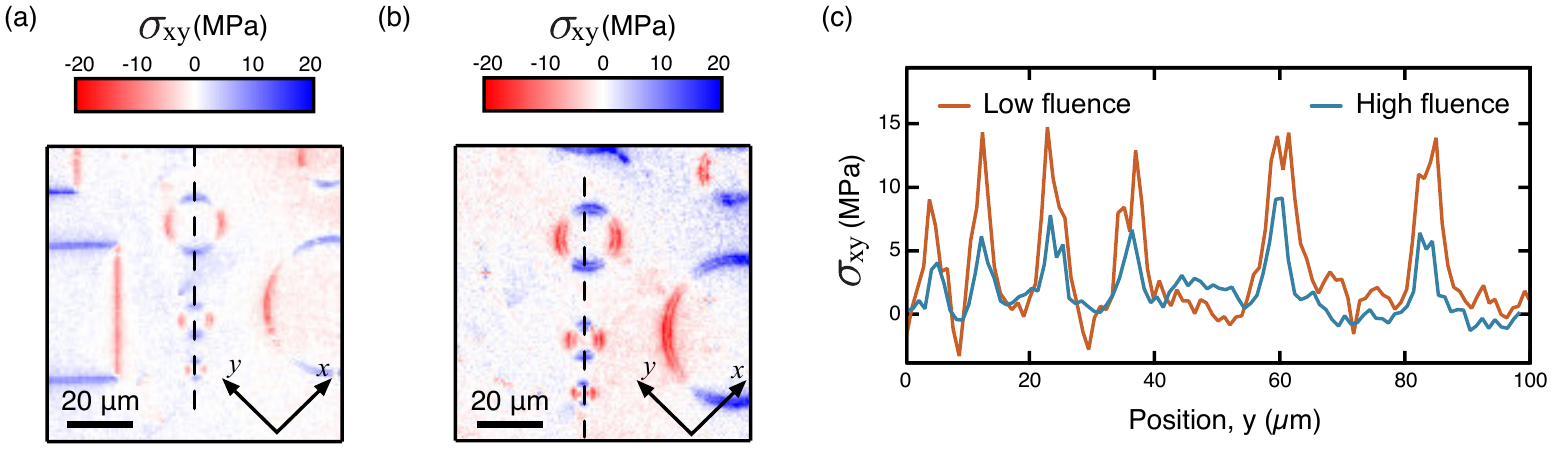}
	\caption{Effect of carbon implant fluence.
		(a,b) $\sigma_{xy}$ map of diamond \#6 (a) and diamond \#2 (b), which were implanted with C$_3$ to a fluence of $3.4\times10^{15}$ and $3.4\times10^{14}$~molecules~cm$^{-2}$, respectively. 
		(c) Line cuts along the dashed lines in (a,b). 
	}
	\label{Fig: dose}
\end{figure}

The stress introduced by the C implant was gauged by recording shifts in the first order Raman peak at 1332 cm$^{-1}$~\cite{Olivero2013,Kato2012, Biktagirov2017}. The volume of the lattice probed by Raman is defined by the 100x objective (NA = 0.95) point spread function rather than, as in the case of NV sensing, the position of the quantum probes and their own range of view. This volume thus encompasses the region of the implant as well as the underlying substrate where a range of strains may exist that act to broaden the Raman signal and limit its accuracy. For simplicity, we assume a hydrostatic stress, $\sigma_{\rm h}=\sigma_{xx}=\sigma_{yy}=\sigma_{zz}$, which can be linked to the change in the Raman shift position, $\Delta\omega$, via the empirical relation $\sigma_{\rm h}=\Delta\omega/\alpha_{\rm h}$, where $\alpha_{\rm h}=2.83$ cm$^{-1}$~GPa$^{-1}$~\cite{Prawer2004}. The Raman measurement was performed with a Renishaw InVia Raman system equipped with a 2400 l/mm grating and a 532~nm wavelength laser which was scanned across the surface of the diamond to generate an image. A spatial map of the diamond line shift across a 5-$\mu$m-wide implanted strip in diamond \#6 is shown in Figure~\ref{Fig: implant charactrisation}c. A line cut across the strip indicates a shift of 0.02$\pm 0.005$ cm$^{-1}$ (Figure~\ref{Fig: implant charactrisation}d). This leads to a stress of $\sigma_{\rm h}=7 \pm 2$~MPa, reasonably consistent with the NV measurements which indicated a maximum axial stress of about 10 MPa in this sample.     

\begin{figure}[t!]
	\centering
	\includegraphics[width=0.9\columnwidth]{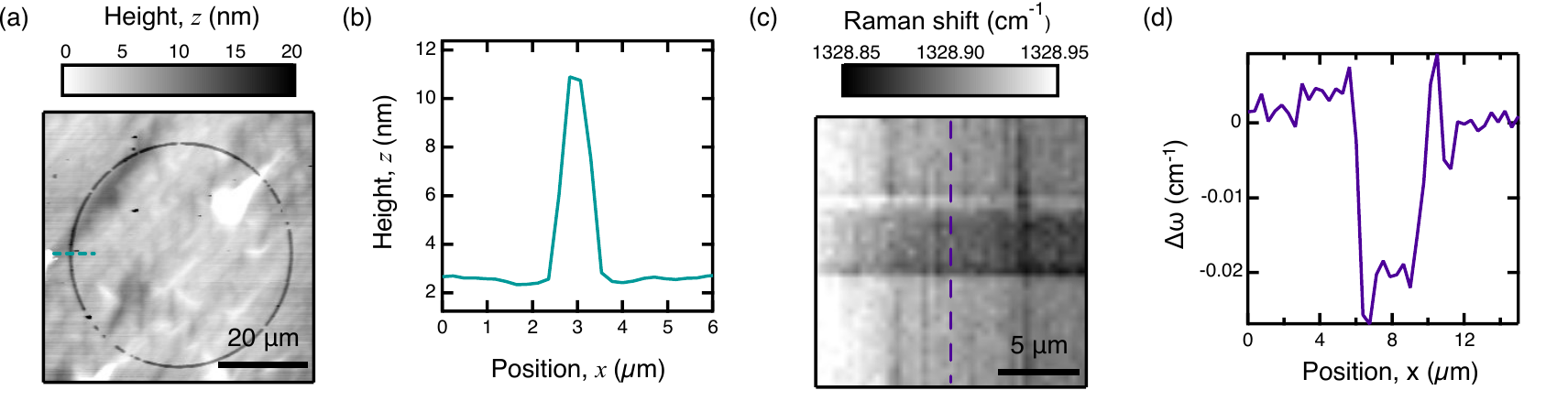}
	\caption{AFM imaging and Raman spectroscopy in diamond \#6. 
		(a) Atomic force microscopy (AFM) image of a ring implant in diamond \#6. The discontinuities in the ring are due to imperfections in the implantation mask.
		(b) Line cut of the AFM image along the dashed line in (a).
		(c) Map of the diamond Raman line in a 5-$\mu$m-wide implanted strip (horizontal). 
		(d) Line cut of the relative shift in the diamond Raman line along the dashed line in (c).
	}
	\label{Fig: implant charactrisation}
\end{figure}

\begin{figure}[t!]
	\centering
	\includegraphics[width=0.9\columnwidth]{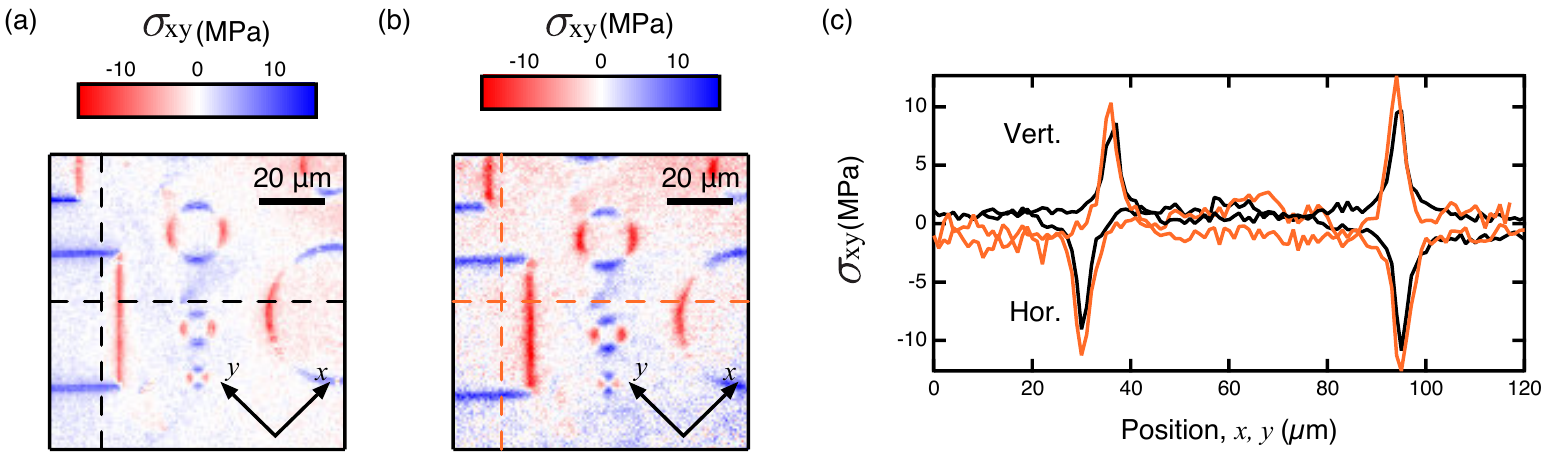}
	\caption{Effect of annealing in diamond \#6.
		(a) $\sigma_{xy}$ map of an implanted region in diamond \#6, before annealing. 
		(b) $\sigma_{xy}$ map of the same region after annealing at 800$^\circ$C for 1 hour. 
		(c) Line cuts along the colour coded dashed lines in (a) and (b). 
	}
	\label{Fig: Annealing}
\end{figure}

We also looked at the effect of annealing the sample, which is expected to convert the amorphous carbon region generated through the implant into graphite. As the graphite has a longer bond length than both the diamond and the amorphous carbon, this procedure will enhance the stress that is embedded. The stress maps ($\sigma_{xy}$) of the same region before and after annealing the sample at 800$^\circ$C for 1 hour are shown in Figure~\ref{Fig: Annealing}a and \ref{Fig: Annealing}b, respectively. The comparison (Figure \ref{Fig: Annealing}c) indicates that there is a small increase in the stress upon annealing of the order of 20\%, similar to previous results~\cite{Olivero2013}. 

\section{Supplementary data: Indents} \label{Sec: Scratches}

The impressions such as the one studied in Figure 3 of the main text were made by nano-indentation using a Berkovich tip. Figure~\ref{Fig: Indents}a shows a typical load-unload curve, indicating a continuous deformation up to the maximum load of P$_{\rm max}=1$~N. The maximum depth beneath the free surface is approximately 130 nm although this could be affected by simultaneous deformation of the Berkovich indenter. The depth of the residual impression was 15 nm in fair agreement with the AFM results. Similar load-unload curves were obtained for the ten indents we performed in the same diamond. Figure~\ref{Fig: Indents}b shows AFM images of three different indents. From these images, we can get a rough estimate of the contact area of the indent, $A\sim 5~\mu$m$^2$, from which we derive the hardness $H_V=\frac{{\rm P}_{\rm max}}{A}\sim200$~GPa. This is larger than the values typically reported for diamond which fall in the range 80-120~GPa~\cite{Novikov1996,Dub2017,Liu2017}, but this can be explained by the indentation size effect arising from the relatively small load used in our experiments~\cite{Grigorev1982,MILMAN2011}. We also note that the sub-surface defects and damage introduced by the N implant (used to create the NVs) may influence the mechanical response of our diamonds.    

\begin{figure*}[h!]
	\centering
	\includegraphics[width=0.8\linewidth]{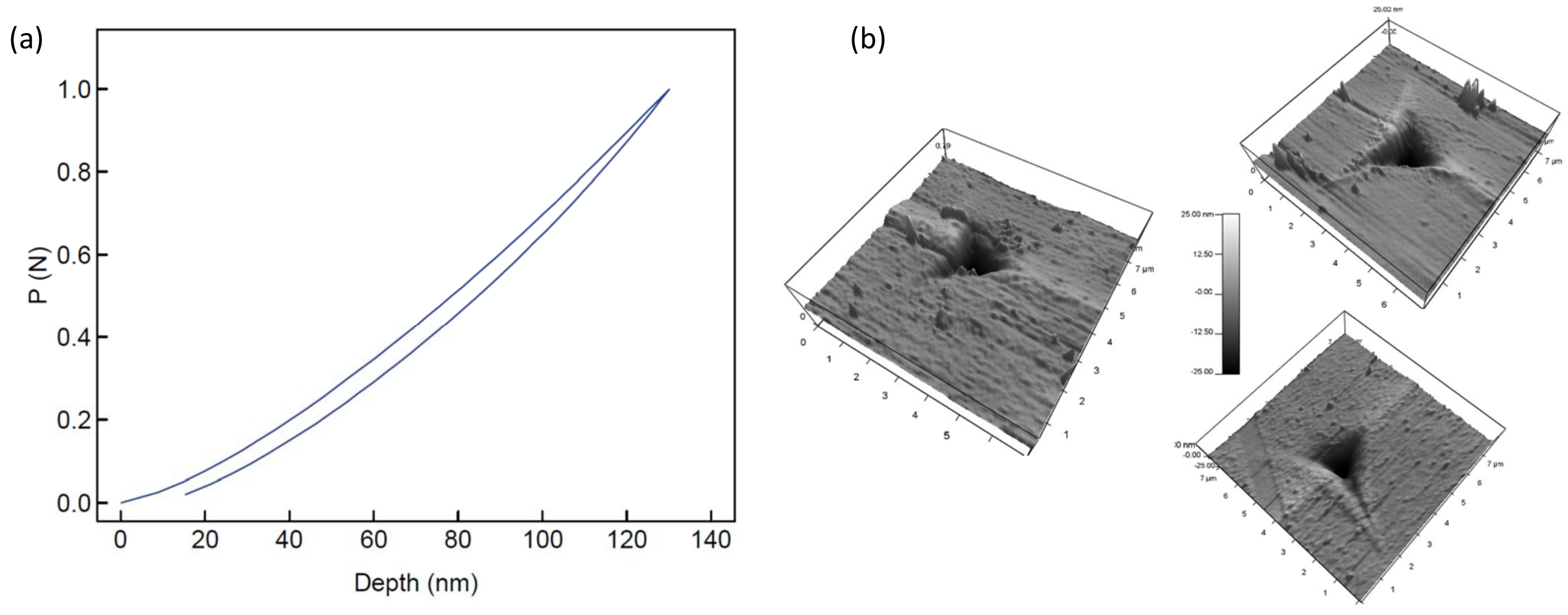}
	\caption{Characterisation of indents in diamond \#3.
		(a) Typical load-unload curve. 
		(b) AFM images of three different indents.
	}
	\label{Fig: Indents}
\end{figure*}

\section{Supplementary data: Scratches} \label{Sec: Scratches}

In Figure 4 of the main text, we showed the body force of a shallow scratch (5 nm deep recess at most) derived from the stress tensor. The full stress maps are shown in Figure~\ref{Fig: Scratches}c, with the PL image recalled in Figure~\ref{Fig: Scratches}a. Furthermore, in Figure 1 of the main text, we showed ODMR spectra of selected pixels near deeper scratches (up to 20 nm deep recesses). The PL image of the corresponding region is shown in Figure~\ref{Fig: Scratches}b, with the stress maps shown in Figure~\ref{Fig: Scratches}d. Here, the stress can be in excess of 1 GPa, which causes some artefacts in the images (and apparent noise in these regions) because the ODMR lines may overlap~\cite{Tetienne2018b}. 

\begin{figure*}[t!]
	\centering
	\includegraphics[width=0.99\linewidth]{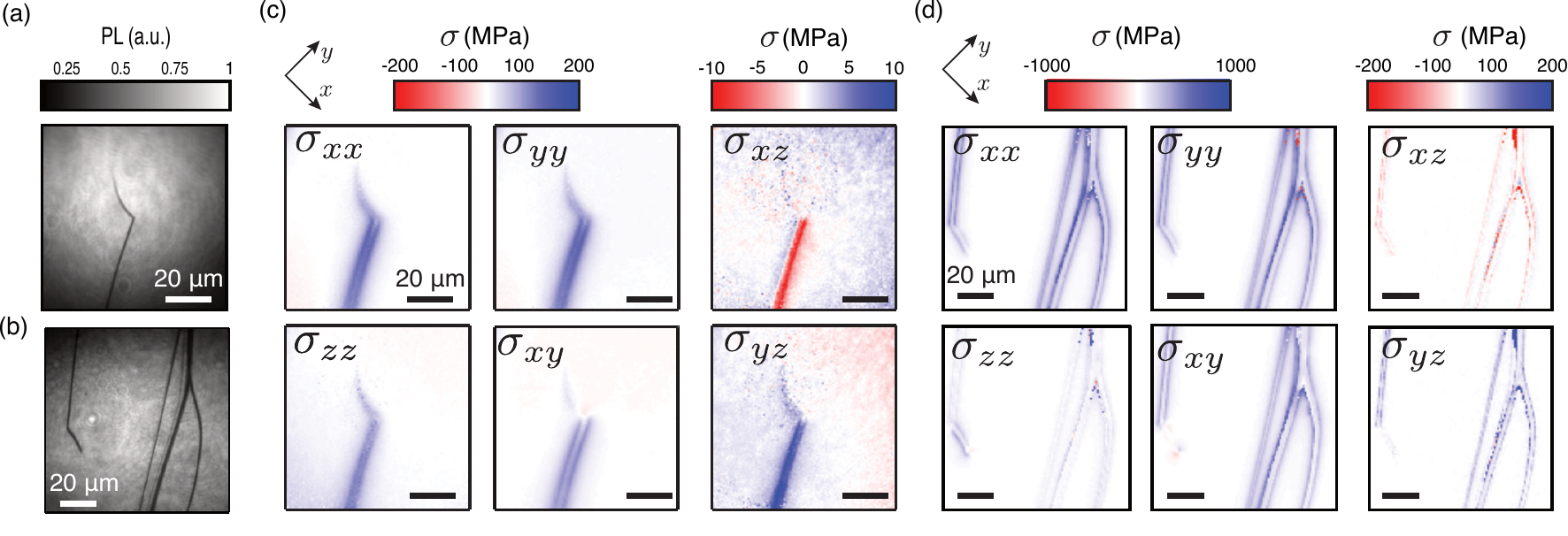}
	\caption{Stress maps of scratches in diamond \#1.
		(a,b) PL images of a region comprising a shallow scratch (a, 5 nm deep) and deep scratches (b, 20 nm deep).
		(c) Maps of the six stress components corresponding to (a).
		(d) Maps of the six stress components corresponding to (b).
	}
	\label{Fig: Scratches}
\end{figure*}       

\end{widetext}

\bibliographystyle{naturemag}
\bibliography{bib}

\end{document}